\documentclass[reprint,amsmath,amssymb,superscriptaddress,twocolumn,aps,prb]{revtex4-2}

\usepackage[utf8]{inputenc}
\usepackage{tikz-cd}
\usepackage{color}
\usepackage{dsfont}
\usepackage{bm}
\usepackage{mathtools}
\usepackage{amsthm}
\usepackage{amsfonts,amssymb,amsmath}
\usepackage{physics}
\usepackage{multirow}
\usepackage{graphicx}

\usepackage{float}
\usepackage{verbatim}
\usepackage{xurl}
\usepackage[hidelinks]{hyperref}

\begin{document}
\title{
Geometric quantum drives and topological dynamical responses:
hyperbolically-driven quantum systems and beyond
}

\author{Jihong Wu}
\affiliation{Department of Physics, National University of Singapore, Singapore 117551}

\author{Chuan Liu}
\affiliation{Department of Physics, National University of Singapore, Singapore 117551}

\author{Daniel Bulmash}
\affiliation{Department of Physics, United States Naval Academy, Annapolis MD 21402 USA}

\author{Wen Wei Ho}
\email{wenweiho@nus.edu.sg}
\affiliation{Department of Physics, National University of Singapore, Singapore 117551}
\affiliation{Centre for Quantum Technologies, National University of Singapore, 3 Science Drive 2, Singapore 117543}

\begin{abstract}
We introduce a geometrical framework to construct a large class of time-dependent quantum systems, in which the position of a classical particle moving autonomously on a smooth connected manifold is used to steer a quantum Hamiltonian over time. This results in quantum drives with structured temporal profiles and properties dependent on the local and global nature of the underlying choice of manifold. We show that our construction recovers the well-known classes of periodically-driven and quasiperiodically-driven quantum systems, but also unveils fundamentally new classes of quantum dynamics: by utilizing a compact 2d hyperbolic Bolza surface and a nonorientable Klein-bottle surface, we demonstrate examples of a hyperbolically-driven quantum system and a nonorientably-driven quantum system respectively. Furthermore, we demonstrate that these driven systems exhibit unusual quantized dynamical responses reflecting their different underlying topologies, under the condition of being fully gapped and in the adiabatic limit, and which have interpretations as quantized crystalline electromagnetic responses in certain exotic effective tight-binding lattice models. We envision geometric quantum driving as a general framework to chart the landscape of time-dependent quantum systems and investigate the universal phase structures they exhibit, as well as a useful tool to enhance the capabilities of modern day quantum simulators. 
\end{abstract}
\maketitle

\section{Introduction}
Time-dependent controls are general and powerful tools in physics, routinely used to engineer new interactions or stabilize states not possible at equilibrium. A classic and well-known example is the Kapitza pendulum, where an originally unstable inverted equilibrium point can be made stable through rapid driving of the pivot~\cite{Kapitsa:1951}. In  quantum systems, driving allows for the control of effective Hamiltonians for cold atoms in optical lattices and quantum materials~\cite{bukov2015universal,eckardt17atomic,oka2019floquet,weitenberg2021tailoring}. In particular, periodic driving has been used to engineer artificial gauge fields and Floquet topological Bloch bands~\cite{kitagawa2010topological, Aidelsburger2015, rudner2020band}. Novel nonequilibrium interacting phases of matter such as the discrete time-crystal~\cite{ PhysRevLett.116.250401, PhysRevLett.117.090402, PhysRevLett.118.030401, annurev:/content/journals/10.1146/annurev-conmatphys-031119-050658} or the anomalous Floquet  insulator \cite{titum2016anomalous, PhysRevB.99.195133} have also been uncovered, which  cannot intrinsically exist without explicit time-dependent modulation of the system.

Out of all possible driven quantum systems, time-periodic or Floquet systems~\cite{bukov2015universal, else2016classification, eckardt17atomic, oka2019floquet, HO2023169297} (and their generalization, time-quasiperiodic systems~\cite{VerdenyPuigMintert+2016+897+907, PhysRevB.98.220509, PhysRevB.99.064306, PhysRevX.10.021032}), wherein system parameters are modulated with a single (few multiple) independent fundamental frequency (frequencies), are perhaps the most well-studied. 
This stems from the conceptual simplicity in understanding the effect of the drive: each frequency acts as an additional synthetic dimension~\cite{PhysRevA.7.2203}, 
and thus physical properties of a complex higher dimensional system may be emulated in a lower dimensional one. 
However, time-periodic or time-quasiperiodic driving constitute only a very limited subset of all possible time-dependent modulations of a system. 
This naturally raises a simple but important conceptual question: what are more general classes of quantum driving $H(t)$, 
generated by a time-dependent quantum Hamiltonian, which can lead to interesting responses and novel phase structure in dynamics? Crucially, the consideration of such generality is not merely academic but also relevant in practice:   time-dependent controls in modern quantum simulator experiments are often implemented using {\it arbitrary waveform generators} (AWG)~\cite{10.1063/1.4795552, Bluvstein2022, PhysRevLett.130.210403}, which in principle allows for the creation of waveforms with arbitrary temporal profiles (within limitations such as finite bandwidths), thus according a multitude of ways to controllably take a system out of equilibrium. 

\begin{figure}[!t]
    \centering
    \includegraphics[width=1\columnwidth]{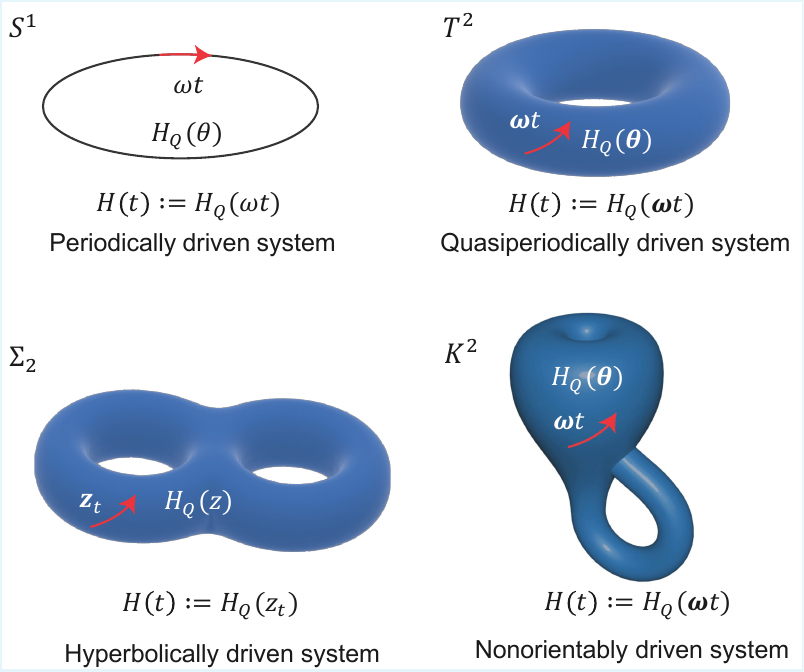}
    \caption{ Geometric quantum drives.  Geodesic flows on a circle $S^1$~(top left),  a torus $T^2$~(top right), a genus-2 torus $\Sigma_2$~(bottom left) and a Klein-bottle surface $K^2$~(bottom right), yield periodically, quasiperiodically, hyperbolically and nonorientably driven quantum systems respectively. In this work, we present explicit constructions of the latter two systems and expound upon the novel physics they exhibit. 
    }
    \label{fig:1}
\end{figure}

A central challenge in tackling this question, though, is that the space of all possible drives  $H(t)$ is very large. Moreover, it can be expected that the vast majority of drives will not lead to any interesting physical phenomena --- for example, a random sequence of pulses applied results in the system responding chaotically and exhibiting a featureless state on average \cite{pilatowsky2023complete}. It is key that one  has to impose {\it structure} on the temporal profiles of the drives in order to achieve nontrivial dynamics \cite{PhysRevLett.116.250401, PhysRevX.7.011026}. Previous works along this front have been scant, but a few notable ones include aperiodic drives generated by sequences of pulses following simple concatenation rules on basic characters of an alphabet (so-called `morphic' or `multipolar' sequences~\cite{PhysRevLett.126.040601,PhysRevB.105.245119,dmfd-lgcq}). These have been predicted to give rise to interesting physical phenomena like time rondeau crystals, a novel form of spatiotemporal ordering~\cite{moon2024experimentalobservationtimerondeau}, and critically-slow exploration of the Hilbert space~\cite{dmfd-lgcq}.

In this work, we make significant headway on this question by introducing a general construction based on geometric considerations, which yields distinct classes of quantum drives $H(t)$ with structured temporal profiles.
%
Our key idea is to use an autonomous classical particle moving on a choice of smooth connected manifold to steer a parent quantum Hamiltonian defined on the manifold; this results in a time-dependent quantum Hamiltonian whose behavior reflects the manifold's local geometry and global topology---motivating the naming of such classes of Hamiltonians as {\it geometric quantum drives}.

Employing the framework, we demonstrate two novel classes of quantum drives, and predict interesting and exotic dynamical topological responses within them. 
Concretely, we first utilize a compact and hyperbolic space called the Bolza surface~\cite{3603e0c0-3eab-3525-9b0a-caac90c781bb} to construct a {\it hyperbolically-driven quantum system}~(HDQS). We show a HDQS exhibits a quantized dynamical response set by the Chern number of its instantaneous bands in the fully gapped phases and adiabatic limit, which can be measured with a specially constructed observable. We then consider nonorientable surfaces like the Klein bottle and the real projective plane, which yield {\it nonorientably-driven quantum systems}~(NDQS). 
We then show that a Klein bottle drive can exhibit a measurable quantized topological dynamical response associated with a dipolar Chern number characterizing the system's topology, and which has the physical interpretation of a crystalline response to an electromagnetic field in an effective
solid-state lattice system with so-called nonsymmorphic momentum-space symmetry~\cite{PhysRevX.14.041060,5nsr-rw4d}.
We envision geometric quantum driving as a powerful framework to chart the landscape of time-dependent quantum systems and investigate the universal phase structures they exhibit, and as well as  a useful tool to extend the simulation capabilities of modern day quantum experiments.

\section{Results}

\subsection{ Geometric quantum driving}
We begin by introducing our theoretical framework of geometric quantum driving. We define a geometric quantum drive to be  specified by a choice of triple $(M,H_Q(x),x_t)$. Here, $M$ is a smooth manifold of dimension $d$, described by local coordinates $x = (x^1,\cdots,x^d)$;  
$H_Q(x)$ is a parent quantum Hamiltonian parametrized by coordinates on $M$, which acts on a $D$-dimensional Hilbert space
(assumed at least piecewise continuous and with bounded norm); and $x_t$ is a trajectory on the manifold. We then construct a time-dependent quantum Hamiltonian as
\begin{align}
H(t):=H_Q(x_t).
\label{eqn:H_from_HQ}
\end{align}
Interpreting $x_t$ as the position of some classical particle moving on $M$, this time-dependent quantum Hamiltonian may be understood as arising from a particle `steering' the underlying parent quantum Hamiltonian (with no back action). 

Stated this way, the construction is rather (perhaps overly!) general --- we have not specified desired properties of the manifold $M$ (like compactness or metricity); nor have we prescribed how the trajectory $x_t$ is to be generated (e.g., through some vector flow etc.), all of which can potentially affect the nature of the resulting driven quantum system $H(t)$. We now impose constraints on the triple $(M,H_Q(x),x_t)$ to make concrete predictions, but we stress that the framework is versatile and expect that different assumptions can lead to different classes of driven quantum systems.


In what follows we restrict ourselves to the case when the manifolds in consideration are closed, i.e., compact and boundaryless; and Riemannian, i.e., there exists a a smooth positive-definite inner product $g_{ij}(x)$ defined on the tangent space at every point $x$. The existence of a Riemann metric allows for the consideration of a natural trajectory $x_t$ on the manifold $M$ --- {\it geodesics} \cite{geodesics}, which we also assume henceforth. 


It is helpful at this stage to  see our abstract discussion in action with these choices: they are powerful enough to already recover the well-known classes of periodically and quasiperiodically driven quantum systems [see Fig.~\ref{fig:1}]. 
Take $M = S^1$, the circle, and use standard coordinates  $\theta \in [0,2\pi]$ (with end-points identified). Consider a bounded piecewise continuous parent quantum Hamiltonian $H_Q(\theta)$ on $S^1$ so it admits a Fourier series representation $H_Q(\theta) = \sum_{n \in \mathbb{Z}} H_n e^{in\theta}$.
The metric on the circle is flat, $g(\theta) = 1$, yielding geodesics  $\theta_t = \omega t + \theta_0 \mod 2\pi$, which describe a particle moving with constant speed or momentum $p_t=\omega$ around the circle. Using Eq.~\eqref{eqn:H_from_HQ}, the resulting time-dependent quantum Hamiltonian is then
\begin{align*}
    H(t) := H_Q(\omega t+\theta_0) = \sum_{n \in \mathbb{Z}}H_n e^{in (\omega t + \theta_0)},
\end{align*}
which is precisely a time-periodic quantum Hamiltonian, driven with frequency $\omega$. 
Similarly, by choosing $M = T^d$, the   $d$-torus for $d \geq 2$, described by local coordinates  $\bm{\theta} = (\theta^1,\cdots,\theta^d) \in [0,2\pi]^d$ with flat metric $g_{ij}(\bm\theta) = \delta_{ij}$, geodesics  are straight lines $\bm{\theta}_t = \bm{\omega} t+\bm{\theta}_0 \mod \bm{2\pi}$ where $\bm{\omega} = (\omega^1,\cdots,\omega^d)$ is a fixed momentum vector. Evaluation of Eq.~\eqref{eqn:H_from_HQ} for a parent quantum Hamiltonian $H_Q(\bm\theta)$ defined on the torus, assuming $\{\omega^i\}$ are rationally independent, gives
\begin{align*}
    H(t):=H_Q(\bm{\omega}t+\bm\theta_0) = \sum_{\bm{n} \in \mathbb{Z}^d}H_{\bm{n}} e^{i\bm{n}\cdot (\bm\omega t + \bm\theta_0)},
\end{align*}
precisely a quasiperiodically-driven quantum Hamiltonian with frequency vector $\bm\omega$.  

Next, we will go beyond these basic and familiar examples of the flat circle $S^1$ and flat torus $T^d$, and show how our geometric quantum driving framework can predict new and distinct classes of quantum drives with interesting physics. We consider two-dimensional manifolds with nontrivial topology, exemplified by a hyperbolic surface and a nonorientable surface.

\begin{figure}[!t]
    \centering
    \includegraphics[width=1\linewidth]{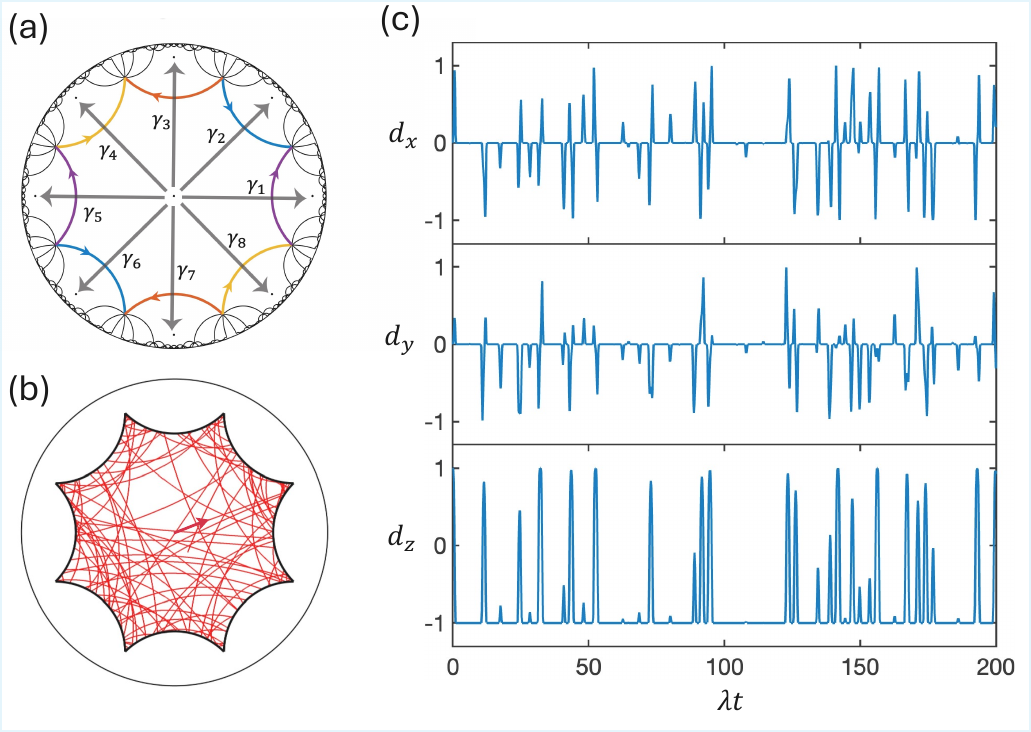}
    \caption{(a) We take $\Sigma_2$ to be the Bolza surface, a compact  Riemann surface with constant negative curvature.
    On the Poincair\'e disk,  the  fundamental domain is an octagon, with opposite sides identified in the orientation shown via Fuchsian translates $\gamma_i$.  
    (b) Geodesic on the Bolza surface.
    (c) Temporal profile of the example hyperbolically driven qubit Hamiltonian  considered in this work with $\epsilon = 0.5$.}
    \label{fig:2}
\end{figure}
\subsection{ Hyperbolically driven quantum system (HDQS)}
Let us  consider a 2d manifold with nontrivial negative curvature: a double torus $\Sigma_2$, see Fig.~\ref{fig:1}.
To be explicit, we take $\Sigma_2$  to  be the Bolza surface, a particular compact, closed, hyperbolic Riemann surface of genus 2~\cite{riemann}, constructed as the quotient space $\mathbb{H}\backslash \Gamma$, where $\mathbb{H}$ is the hyperbolic plane  and $\Gamma$ is a Fuchsian group of translates; see \cite{fuchsian} and the Supplemental Information~(SI) for details~\cite{SM}. In the Poincair\'{e} disk model $|z|<1$ which has  metric $g_{ij}(z) = g(z)\delta_{ij}$ with $g(z) = 4/(1-|z|^2)^2$, the fundamental domain is an octagon wherein opposite sides (which are arcs of circles) are identified with orientation shown in Fig.~\ref{fig:2}. 
We note the Bolza surface is not only a mathematical construct but has featured in various areas of physics: the Hadamard-Gutzwiller model~\cite{Hadamard1898,PhysRevLett.45.150, gutzwiller1985geometry,PhysRevLett.61.483} describes  the motion of a free particle on the surface (an example of dynamical billiards), and is a minimal model of classical and quantum chaos;
more recently, the Bolza surface has appeared  in hyperbolic band theory~\cite{doi:10.1126/sciadv.abe9170} as a unit cell involved in a particular tiling of hyperbolic space. Our work adds to the growing list of interesting physical phenomena tied to hyperbolic geometry.

To understand  the nature of geodesics on the Bolza surface, it is helpful to adopt the viewpoint of geodesics as arising from Hamiltonian dynamics of a free classical particle on the surface~\cite{geodesics}. Precisely, let $H_{cl}(z,p) = \frac{1}{2} g^{ij}(z) p_ip_j$ be the classical Hamiltonian describing a unit-mass free classical particle on $\Sigma_2$. This is a function on   phase space described locally by position and momentum coordinates $(z,p):=(x^1+ix^2,p_1+ip_2)\in \mathbb{C}^2$; more precisely, the phase space is the  cotangent bundle $T^*\Sigma_2$ of the Bolza. Here, $g^{ij}(z)=g(z)^{-1}\delta_{ij}$ is the inverse metric. Then,  Hamiltonian's equations yield~\cite{SM}
\begin{align}
\dot{z} &= \frac{1}{4}(1-|z|^2)^2p, \qquad 
\dot{p} = \frac{1}{2} z(1-|z|^2)|p|^2.
\end{align} 
Upon specifying an initial condition $(z_0,p_0)$, there is a unique solution $(z_t,p_t)$;
the resulting $z_t$ is the desired geodesic trajectory~(note an appropriate Fuchsian translate has to be applied whenever the particle exits the fundamental region, see Fig.~\ref{fig:2}(b) and \cite{SM} for details). 
Note that given a geodesic solution $(z_t,p_t)$, we can always construct an infinite family of related geodesic solutions 
$  
(z^{(\lambda)}_t, p^{(\lambda)}_t) := (z_{\lambda t}, \lambda p_{\lambda t})
$  
for any $\lambda>0$, each of which describes the same {\it path} on the manifold $\Sigma_2$ but differs by its conserved energy 
\begin{align*}
    H_{cl}(z_{\lambda t},\lambda p_{\lambda t}) = H_{cl}(z_{\lambda 0},\lambda p_{\lambda 0}) = \frac{1}{2}g(z_{0})^{-1}|p_{0}|^2 \lambda^2;
\end{align*}
equivalently, they are geodesics of different speeds $\lambda$. We will henceforth assume that the normalization is such that $\lambda=1$ corresponds to the unit-speed geodesic with energy $1/2$.

Lastly, given a choice of parent quantum Hamiltonian $H_Q(z)$ defined on the Bolza surface, we define a {\it hyperbolically driven quantum system} (HDQS) as
\begin{align}
H_{\lambda}(t):=H_Q(z^{(\lambda)}_t),
\end{align}
where $\lambda$ tunes the `speed' that the parent quantum Hamiltonian is being driven at, in analogy to the frequency  $\omega$ in a periodically-driven system. By virtue of our geometric construction, this is a class of quantum drives with temporal profiles clearly distinct from time-periodic or time-quasiperiodic ones, see Fig.~\ref{fig:2}(c) for an example given by the explicit qubit model discussed later.

While our construction has given rise to a Hamiltonian with exotic temporal profile, we now explain that HDQS can exhibit novel physical phenomena. 
We will show how HDQS which have the property of being `fully gapped' (defined below),  can be completely topologically
characterized by  quantized dynamical responses in the adiabatic limit, which are related to the underlying topology of the Bolza surface $\Sigma_2$.

Let us call a $D$-level smooth parent quantum Hamiltonian $H_Q(z)$  `fully gapped'  if its eigenstates $|\psi_n(z)\rangle$  have energies $E_n(z)$ obeying   $E_n(z) < E_{n+1}(z)$ point-wise for every $z \in\Sigma_2$. Thus each  $|\psi_n(z)\rangle$ ($n = 1,\cdots, D$) forms a well-separated `band' over the manifold $\Sigma_2$, the latter of which can be  interpreted as a generalized Brillouin zone (BZ) in analogy to band theory of solid state physics (though our BZ crucially did {\it not} arise as the momentum space description of some lattice in either Euclidean or hyperbolic space). The topological properties of quantum systems parametrized by coordinates on a generic manifold $M$ have been studied in great generality recently~\cite{KitaevTalk,KapustinBerry,KapustinChargePump,BeaudryParametrized,RyuHigherStructures}, but in our present case, standard arguments tell us that the topological classification of such bands is given by the obstruction to 
choosing a globally smooth gauge for $|\psi_n(z)\rangle$ over the surface~\cite{bernevig}. This is captured by the $\mathbb{Z}$-valued first Chern number 
\begin{align}
C_1^{(n)} =\frac{1}{2\pi}\iint_{\Sigma_2}dx^1dx^2
\Omega^{(n)}_{x^1 x^2}(z),
\end{align}
 the integral of the Berry curvature form $\Omega^{(n)}_{x^1 x^2}(z)=\partial_{x^2}A^{(n)}_{x^1}(z)-\partial_{x^1}A^{(n)}_{x^2}(z)$ over the fundamental region in the Poincair\'e disk defining the Bolza surface, where $A^{(n)}_{x^j}(z)= i\langle \psi_n(z)|\partial_{x^j}\psi_n(z)\rangle$ is the Berry connection~(note: there is no Poincair\'e metric under the integral!). 
 More sophisticatedly, the first Chern number can be understood as a   topological classification of $U(1)$ fiber bundles on a genus 2 Riemannian surface (here, the Bolza surface), and takes values in the second cohomology group $H^2(\Sigma_2,\mathbb{Z}) = \mathbb{Z}$~\cite{characteristic_class}. 

Intriguingly, we claim the (global) topological invariant  $C_1^{(n)}$ characterizing the $n$-th band 
may be extracted in the  {\it dynamics} of the (local) $D$-level HDQS, as follows: fixing some choice of geodesic $(z_t^{(\lambda)},p_t^{(\lambda)})$, 
initialize the quantum system in the $n$th instantaneous eigenstate $|\psi_n(z_0^{(\lambda)})\rangle$ at $t=0$, and evolve the system with  $H_\lambda(t):=H_Q(z^{(\lambda)}_t)$ to obtain $|\psi(t)\rangle$. 
Then, measure the following judiciously-constructed time-dependent observable of the $D$-level quantum system:
\begin{align}
O^{(\lambda)}(t):= 2g(z_t^{(\lambda)})^{-1}\left(p^{(\lambda)}_{2,t} \partial_{x^1}H_Q(z^{(\lambda)}_{t})-1\leftrightarrow2\right).
\label{eqn:Wt}
    \end{align}  
While the form of $O^{(\lambda)}(t)$ looks complicated, it is simply a observable constructed from the parent quantum Hamiltonian and thus has a  time-dependence explicitly  determined by the choice of geodesic which can be easily written down. 
Finally, evaluate its (scaled) time-average expected value 
\begin{align}
w(T) := \frac{1}{\lambda^2 T}\int_0^Tdt  \langle \psi(t)|O^{(\lambda)}(t)|\psi(t)\rangle.
\end{align} 

\begin{figure*}[!t]
    \centering
    \includegraphics[width=1\linewidth]{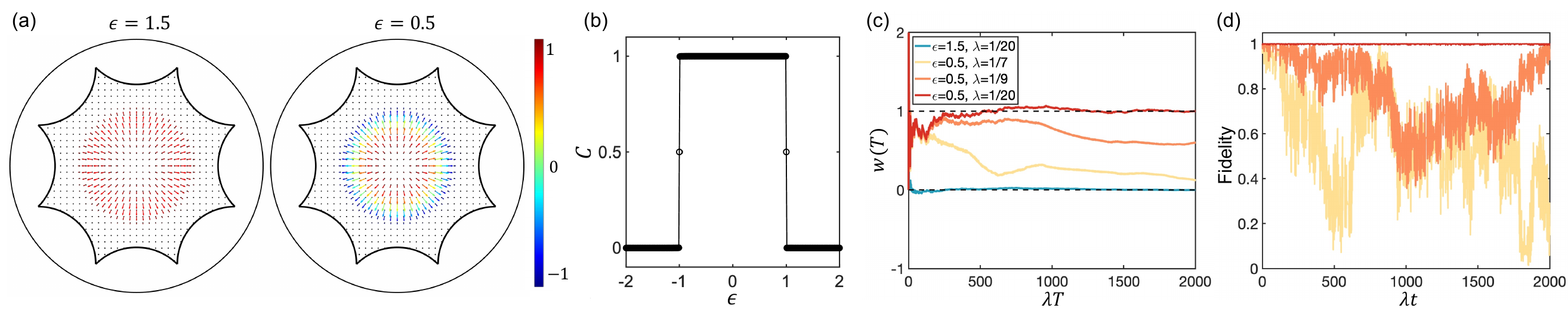}
    \caption{(a) Spin texture $\bm{d}_\epsilon(z)$ of the parent qubit Hamiltonian $H_Q(z)$ on the Bolza surface for   $\epsilon = 1.5, 0.5$. When $\epsilon = 1.5(0.5)$, the texture is ferromagnetic(skyrmionic), pointing mostly up(inverting from up to down).
    (b) The difference in textures can be quantitatively and sharply captured by the Chern number of the upper band of $H_Q(z)$, which changes from $0$ (trivial) when  $|\epsilon|>1$, to $1$ (topological) when $|\epsilon|<1$. (c) Dynamical response $w(T)$ beginning from the instantaneous eigenstate of the upper band. For small $\lambda$, $w(T)$ converges to the Chern number of the band at late times, in both trivial and topological regimes. (d) Fidelity between time evolving state and instantaneous eigenstate for different $\lambda$s and $\epsilon=0.5$.
    }
    \label{fig:4}
\end{figure*}

We argue that in the long-time and slow-driving limit (taken in that order), the response $w(T)$ is topologically quantized to the first  Chern number  for almost all choices of  geodesic flows~\cite{SM}:
\begin{align}
\lim_{\lambda \to 0} \lim_{T \to \infty} w(T) = C_1^{(n)}. 
\label{eqn:claim}
\end{align}
Eq.~\eqref{eqn:claim}, the topologically quantized dynamical response of a HDQS, together with the identification of the nontrivial obserable Eq.~\eqref{eqn:Wt}, constitute our main results in this section. In practice, a large $T$ and small $\lambda$ (but with both being finite) is sufficient to see good quantization, as we will see with the explicit model below [Fig.~\ref{fig:4}(c)]. 
We note that perfect quantization in a fully-gapped HDQS can be achieved without the need for slow driving, with the use of counterdiabatic driving, see SI for details~\cite{Berry_2009,doi:10.1073/pnas.1619826114,SM}.

We now briefly explain why one should expect a quantized topological response.  We argue that the quantization relies on two key properties of these HDQS which generalize to many other choices of $M$ and many other choices of trajectories. First, under slow driving (more precisely, when the speed $\lambda$ is much less than the energy gaps between the $n$th and $n\pm1$-th `bands' --- which can always hold due to the fully-gapped condition), the system stays close to the $n$th instantaneous band by the quantum adiabatic theorem~\cite{Born1928-hv}. That is, $|\psi(t)\rangle \approx |\psi_n(z_t^{(\lambda)})\rangle + \cdots $ (we have omitted the standard geometric and dynamical phases accumulated), where $\cdots$ represent corrections in $\lambda$ in an adiabatic expansion~\cite{adiabatic}. Thus, $w(T)$ may also be expanded in an adiabatic series. Second, our choice of path (gedoesic) is \textit{ergodic} on the Bolza surface $\Sigma_2$. This nontrivial fact is guaranteed by a theorem of Hopf's
\cite{hopf1936fuchsian,bams/1183533160}, 
which states more generally that 
{\it
the geodesic flow on a closed, negatively curved surface is ergodic}.
For the Bolza surface, the implication (invoking Birkhoff's theorem~\cite{doi:10.1073/pnas.17.2.656}) is then as follows: 
consider an integrable function $f(z,p)$ defined on the cotangent bundle $T^*\Sigma_2$ of the Bolza surface, and its time-average along some geodesic $(z^{(\lambda)}_t, p^{(\lambda)}_t)$. For almost all initial conditions $(z_0^{(\lambda)},p_0^{(\lambda)})$, as $T \to \infty$,
\begin{align*}
\int_0^T \!\!\frac{dt}{T} f(z_t^{(\lambda)},p_t^{(\lambda)}) \to \iint_{\Sigma_2} \!\!\!\!\! dx^1 dx^2\frac{g(z)}{4\pi} \int_0^{2\pi} \!\!\! \frac{d\theta}{2\pi}f(z,\lambda g(z)^{\frac{1}{2}}e^{i\theta}), 
\end{align*}  
(see Supplementary Figure 2 in SI \cite{SM} for illustration).
The meaning of the RHS is as such: fix a point $z$ and consider the possible momenta of the classical particle  whenever it passes through that point. By energy conservation, these all have fixed length $\lambda g(z)^{\frac{1}{2}}$; so the only remaining degree of freedom is their direction $e^{i\theta}$. Ergodicity of the motion says that over long times, $\theta$ is sampled uniformly (over the circle). Further, it says this behavior occurs uniformly (with respect to the Poincair\'e metric) over the surface $\Sigma_2$; together this is nothing more than the familiar idea of ergodicity as time-averaging equaling space-averaging (over the slice of $T^*\Sigma_2$ with constant energy $E = \lambda^2/2$).
Utilizing this, we show in the SI~\cite{SM} that the time-averaged observable $w(T)$ at large $T$ can be approximated by a space-integral over the Bolza surface; evaluating this  the leading order term in $\lambda$ vanishes, while the next order term is proportional to the Chern number $C_1^{(n)}$, leading to our claim Eq.~\eqref{eqn:claim}.
Importantly, it is clear that our result of  quantized dynamical responses Eq.~\eqref{eqn:claim} in fact  constitute a  topological {\it characterization} of fully-gapped HDQS: their quantized responses are robust under perturbations of the  system which preserve their fully-gapped nature.
We expect the responses are robust too under  perturbations of the underlying classical path (e.g.,~noise) on $\Sigma_2$, which determines the temporal profile of the Hamiltonian $H(t)$, as long as they are ergodicity-preserving --- we leave these investigations for future work.

To substantiate our analytical claims, we explicitly construct an example qubit HDQS and numerically simulate its dynamical response. Consider the parent quantum Hamiltonian 
\begin{align}
H_Q(z)&=\bm{d}_\epsilon(z)\cdot \bm{\sigma}, 
\label{eqn:Hqubit}
\end{align}
where $\bm\sigma = (\sigma^x,\sigma^y,\sigma^z)$ are standard Pauli matrices and $\bm{d}_\epsilon(z)=(d_{x,\epsilon}(z),d_{y,\epsilon}(z),d_{z,\epsilon}(z))$ is  a  real vector of field strengths varying over $\Sigma_2$, tunable by a parameter $\epsilon \in \mathbb{R}$, defined as follows. 
We introduce a bump function $f(z)$, which changes smoothly and compactly from $\pi/2$ at $z=0$ to $-\pi/2$ at $|z|=0.6$,
whose  precise form is presented in the Methods section (we refer to \cite{SM} for more details).  Then, $\bm{d}_{\epsilon}(z)$ is defined, for $|\epsilon|\neq 1$, as $\bm{d}_{\epsilon}(z) = \bm{d'}_\epsilon(z)/|\bm{d}'_\epsilon(z)|$ where   
\begin{align}
    \boldsymbol{d}'_\epsilon(z)
    =\begin{pmatrix}
        \cos(f(z)) x^1/|z|\\
        \cos(f(z)) x^2/|z|\\
        \sin(f(z)) +\epsilon
    \end{pmatrix};
\end{align}
see Fig.~\ref{fig:4}(a) for illustrative spin textures    $\bm{d}_\epsilon(z)$.
Since  $|\bm{d}_\epsilon(z)|=1$ for every $z$, $H_Q(z)$ is fully gapped. Furthermore, when $\epsilon>1(\epsilon<1)$, $d_{z,\epsilon}$ is always positive(negative), describing a ferromagnetic spin texture, while when $|\epsilon|<1$, the spin texture inverts moving from  the origin  to the boundary, describing a skyrmionic texture. We thus expect the bands of $H_Q(z)$ to be characterized by differing topological numbers in these two parameter regimes (respectively, trivial and topological), which is indeed seen in the Chern number plot, Fig.~\ref{fig:4}(b). 

Turning to quantum dynamics, we consider a family of geodesics with    varying speeds $\lambda$ defined by initial conditions $z_0 = 0$ and $p_0 = \lambda e^{i \pi/9}$ [Fig.~\ref{fig:2}(b)], which defines the hyperbolically driven quantum Hamiltonian $H(t):=H_Q(z_t)$ [Fig.~\ref{fig:2}(c)]. We then measure $O^{(\lambda)}(t)$ and average it to yield $w(T)$.
We see that when $\lambda \lesssim 1/20$, the motion is very close to adiabatic [Fig.~\ref{fig:4}(d)]. Moreover, $w(T)$ converges at late times to the Chern number $0(1)$ in the trivial(topological) regimes of the model [Fig.~\ref{fig:4}(c)], demonstrating its nontrivial topological responses, Eq.~\eqref{eqn:claim}, as predicted.

\subsection{Nonorientably-driven quantum systems (NDQS)}

We next use our geometric quantum driving framework to construct yet another new and distinct class of driven quantum systems derived from the choice of {\it nonorientable} surfaces, i.e.,  2d manifolds in which no globally consistent choice of normal direction can be defined, and unveil the physics it exhibits. In what follows we will focus on the paradigmatic example of the Klein bottle $K^2$, see Fig.~\ref{fig:1}, but in the SI, we also provide another example given by the choice of real projective plane $RP^2$. Collectively, these can be termed {\it nonorientably-driven quantum systems} (NDQS).

\begin{figure}[!t]
    \centering
    \includegraphics[width=1\linewidth]{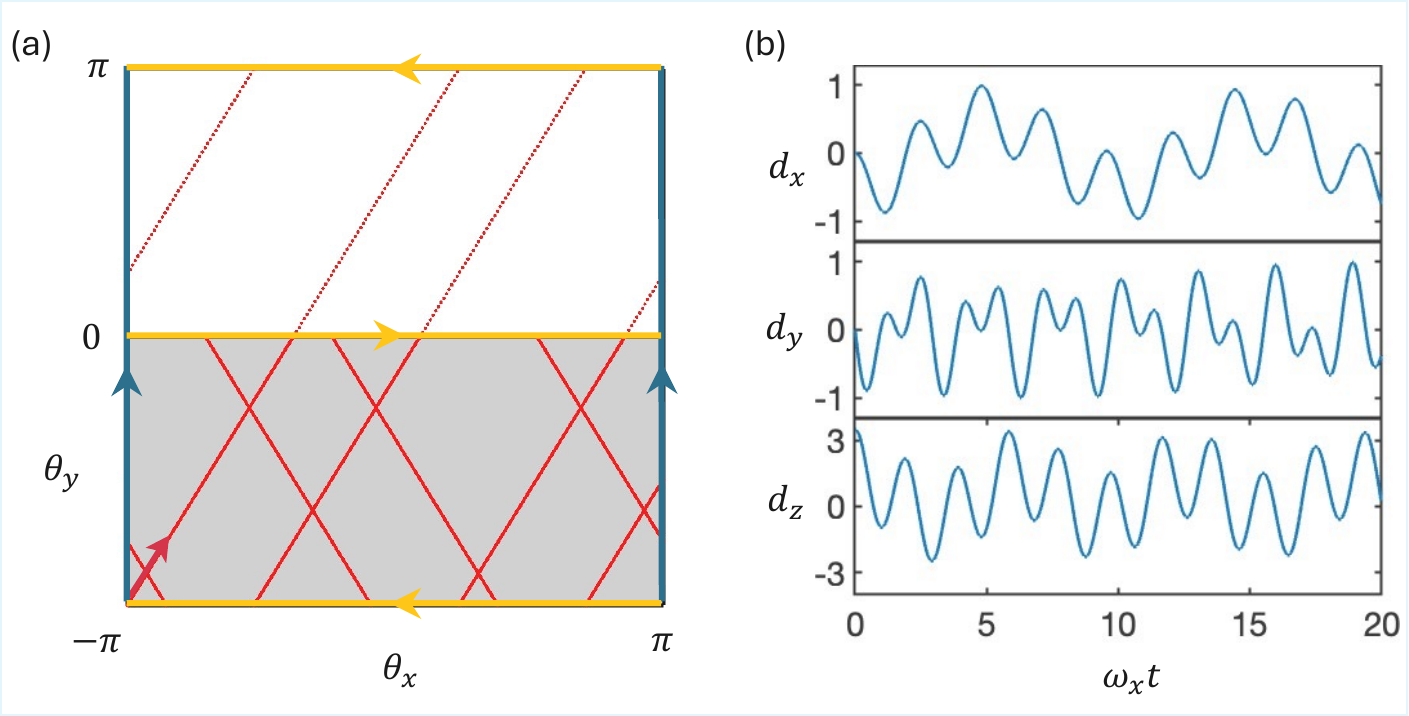}
    \caption{(a) Geodesics on the Klein bottle (gray region) are depicted by red straight lines. Every time the particle leaves the gray region, it gets mapped back to a point within the fundamental region following the identification rules of the boundaries (yellow and blue arrows).  
    Equivalently, one may consider two gluing copies of the Klein bottle as shown (white and gray regions) which yield a standard torus; a Klein bottle geodesic is then equivalent  to a standard geodesic on the torus (dotted trajectory).
    (b) Temporal profile of the example Klein-bottle driven qubit Hamiltonian Eq.~\eqref{eqn:example_NDQS} utilized for numerical simulations, with $m = 0.5$.}
    \label{fig:3}
\end{figure}

Now, the Klein bottle can be realized as the quotient space $\mathbb{R}^2/\tau$, where $\mathbb{R}^2$ is the flat real plane and $\tau$ is the group generated by the following two transformations:
\begin{equation}
    \left\{\begin{aligned}
        \tau_1&:(x,y)\mapsto (x+2\pi,y),\\
        \tau_2&:(x,y)\mapsto(2\pi-x,y+\pi).
    \end{aligned}
    \right.
\end{equation}
As a result, the fundamental domain for $K^2$ can be taken to be $ [-\pi,\pi]\times[-\pi,0] \ni (\theta_x,\theta_y) = \bm{\theta}$, where vertical boundaries are identified with similar orientation 
while horizontal boundaries are identified with opposite orientation (gray region in Fig.~\ref{fig:3}(a)). We note that two Klein bottles can be glued to form a standard torus as illustrated in Fig.~\ref{fig:3}(a) (white and gray regions); indeed,  this stems from the fact that $\tau_1$ and $\tau_2^2:(x,y)\mapsto(x,y+2\pi)$ generate the translation group of the torus.

As the metric on $K^2$ is flat, geodesics on it are generated by trajectories of a free particle, just like for the torus --- with a key difference being how the particle traverses the boundaries, see Fig.~\ref{fig:3}(a) for an example. 
Let us specify the particle's initial velocity $\boldsymbol{\omega}=(\omega_x,\omega_y)$ and initial position $\boldsymbol{\theta}_0=(\theta_{x,0},\theta_{y,0})$. Then, the particle's position in the fundamental domain can be explicitly computed as follows:
\begin{align}
\theta_x(t)& =(-1)^{n_y(t)}[\pi- (\omega_xt+\theta_{x,0}+\pi\mod 2\pi)], \nonumber \\
\theta_y(t)&=(\omega_yt+\theta_{y,0}\mod \pi)-\pi,
\end{align}
where $n_y(t)\equiv\lfloor (\omega_y t+\theta_{y,0})/\pi \rfloor$. These formulae simply encode a trajectory respecting the nontrivial gluing of the boundaries of the fundamental region to form  $K^2$.

Following our framework, we next consider a parent quantum Hamiltonian $H_Q(\bm{\theta})$ defined on the fundamental domain of $K^2$. Combining everything, we finally construct a `Klein bottle drive'  as $H(t):=H_Q(\bm\theta_t)$.
An example temporal profile is shown in Fig.~\ref{fig:3}(b), generated by a concrete qubit model used later in our numerical simulation.
Since two copies of the Klein bottle can be combined to form a torus (as noted above), Klein bottle driving can be thought of as a special subclass of quasiperiodic driving, but with additional structure. 

What sorts of physics do Klein bottle quantum drives exhibit? First we note that because the surface $K^2$ is nonorientable, a conventional first Chern number does not exist characterizing its topology.
Thus,  one cannot hope to find  an observable like Eq.~\eqref{eqn:Wt} yielding a topologically quantized dynamical response like of Eq.~\eqref{eqn:claim} like in the HDQS case.

However, consider now that the parent quantum Hamiltonian $H_Q(\boldsymbol{\theta})$ obeys an additional $y-$mirror symmetry
\begin{align}
    M_y: H_Q(\theta_x,\theta_y)=U^\dagger H_Q(\theta_x,-\theta_y)U,
    \label{eqn:y-mirror_symmetry}
\end{align}
where $U$ is a unitary operator satisfying $U^2=\pm \mathbb{I}$. 
It was recently shown in studies of topological insulators in solid state physics that gapped energy bands $|\psi_n(\boldsymbol{\theta})\rangle$ of such $H_Q(\boldsymbol{\theta})$ {\it are} topologically classified by an invariant quantized to integer multiples of $\pi/2$
~\cite{PhysRevX.14.041060,5nsr-rw4d,DartedChern}:
\begin{align}
    D_y^{(n)} =\frac{1}{2\pi}\iint_{K^2}d\theta_xd\theta_y\theta_y
\Omega^{(n)}(\boldsymbol{\theta}),
\label{eqn:DipolarChern}
\end{align}
called the dipolar Chern number. Above, $\Omega^{(n)}(\boldsymbol{\theta})=\partial_{\theta_y}A^{(n)}_{\theta_x}(z)-\partial_{\theta_x}A^{(n)}_{\theta_y}(\boldsymbol{\theta})$ is the Berry curvature form,  $A^{(n)}_{\theta_j}(\boldsymbol{\theta})= i\langle\psi_n(\boldsymbol{\theta})|\partial_{\theta_j}\psi_n(\boldsymbol{\theta})\rangle$ the Berry connection, and the integral is taken over the fundamental domain of the Klein bottle.
Note in these solid state works \cite{chen2022brillouin, PhysRevX.14.041060,5nsr-rw4d,DartedChern}, the Klein bottle $K^2$ (and also the real projective plane $RP^2$ \cite{5nsr-rw4d}) arose as the Brillouin zone of lattice models with the imposition of certain unusual nonsymmorphic momentum-space symmetries. It was further shown that the dipolar Chern number Eq.~\eqref{eqn:DipolarChern} physically characterizes the (quantized) rate of a momentum current generated in response to an applied electric field, or equivalently the (quantized) charge current induced by time-dependent strains on the lattice.




\begin{figure*}[!t]
    \centering
    \includegraphics[width=1\linewidth]{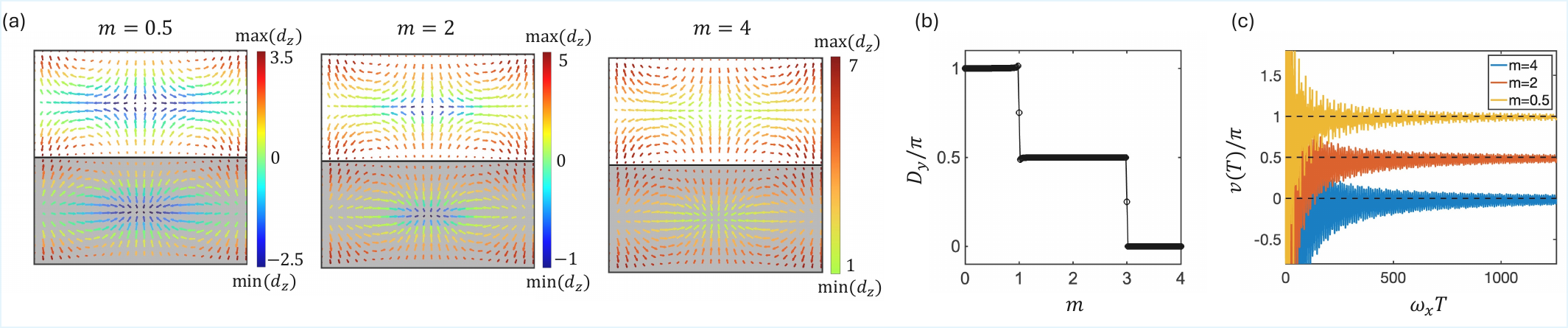}
    \caption{
(a) Spin texture $\bm{d}(\bm\theta)$ defining the parent qubit Hamiltonian $H_Q(\boldsymbol{\theta})$ on the Klein bottle surface (gray regions) for   $m = 0.5, 2,4$ respectively. (b) The corresponding dipolar Chern number versus $m$ is shown. For generic $m$ it is quantized in units of $\pi/2$. 
(c) Dynamical response $v(T)$ starting from the instantaneous eigenstate of the upper band. For the small driving frequency $\bm\omega$ stated in the main text, $v(T)$ approaches the dipolar Chern number of the band at late times. }
    \label{fig:5}
\end{figure*}

In our context,  these exotic manifolds enter simply as   {\it primitive} ingredients in our geometric quantum driving framework, which are simply and directly realized by time-dependent controls.  Moreover, the topological invariant $D_y^{(n)}$ can be measured in our setup as follows: assuming a `fully gapped' condition on $H_Q(\bm\theta)$  (again, this means all its energy levels $|\psi_n(\boldsymbol{\theta})\rangle$ form well-defined `bands' over $K^2$), first 
prepare an instantaneous eigenstate $|\psi_n(\bm\theta_0)\rangle$ at $t = 0$ 
and evolve the system under $H(t):=H_Q(\bm\theta_t)$, where $\bm\theta_t$ describes a geodesic on the Klein bottle. Then measure the following time-dependent quantum operator:
\begin{equation}
    O^{(\boldsymbol{\omega})}(t):=\omega_{y}\theta_{y}(t)\frac{\partial H_Q(\bm\theta_t)}{\partial\theta_x}.
    \label{eqn:O_claim}
\end{equation}
Again, this operator is constructed from the parent quantum Hamiltonian $H_Q(\bm\theta)$ and choice of geodesic $\bm\theta_t$ and thus has a form which can be explicitly determined. 
Finally, we average its expectation value over time:
\begin{equation}
    v(T)=\frac{\pi}{\omega_y^2T}\int_0^Tdt\langle\psi(t)|O^{(\boldsymbol{\omega})}(t)|\psi(t)\rangle.
    \label{eq:non-orient}
\end{equation}
We claim that in the long-time and slow driving limit ($\|\bm\omega \|\ll 1$, keeping the direction of $\bm \omega$ fixed), $\nu(T)$ for generic driving parameters converges corresponding to the dipolar Chern number:
\begin{equation}
    \lim_{\|\bm\omega\|\to0}\lim_{T\to\infty}\nu(T)=D_y^{(n)}.
    \label{eqn:Dn_claim}
\end{equation}
Eqs.~\eqref{eqn:O_claim} and \eqref{eqn:Dn_claim} constitute our main claim in this section --- a nontrivial topological response exhibited by NDQS, whose derivation mirrors that of the HDQS (noting that geodesics on the Klein bottle are ergodic for generic driving frequencies, which allows us to replace the time-integral with the space integral in the long-time limit).
In practice, a clear quantization can already be observed  by taking $T$ large and $\| \bm\omega \|$ small (but both finite), as illustrated in the explicit model of Fig.~\ref{fig:5}.


To illustrate our predictions, we perform numerical simulations of a qubit Klein bottle-driven system, defined by parent quantum Hamiltonian
\begin{equation}
    H_Q(\bm\theta)=\bm d(\bm\theta)\cdot\bm\sigma,
    \label{eqn:example_NDQS}
\end{equation}
and 
\begin{align}
    \boldsymbol{d}(\boldsymbol{\theta})
    =\begin{pmatrix}
        \sin\theta_x\sin\theta_y\\
        \cos\theta_x\sin2\theta_y\\
        m-\cos\theta_x+2\cos2\theta_y
        \label{eq:kleinbottle_model}
    \end{pmatrix},
\end{align}
where $m$ is a tunable parameter, motivated by~\cite{5nsr-rw4d}. This Hamiltonian possesses $y-$mirror symmetry Eq.~\eqref{eqn:y-mirror_symmetry} with the symmetry operator $U=\sigma^z$. Spin textures $\bm{d}(\bm\theta)$ are illustrated in Fig.~\ref{fig:5}(a) for different $m$. It can be shown that the system is fully-gapped for generic values of $m$; in particular, it exhibits topologically nontrivial bands for $0<m<1$ with $|D_y|=\pi$ and for $1<m<3$ with $|D_y|=\pi/2$, while for $m > 3$, they are trivial with $|D_y|=0$, see Fig.~\ref{fig:5}(b). 
For our simulations, we initialize the system at $t = 0$ in the instantaneous upper band at 
$(\theta_x,\theta_y)=(-\pi,-\pi)$, then drive the system with a geodesic with $\bm\omega$  chosen as $\omega_x=0.02$ and $\omega_y=\frac{1+\sqrt{5}}{2}\omega_x$, before measuring $O^{(\bm \omega)}(t)$ and then computing $\nu(T)$.  As illustrated in Fig.~\ref{fig:5}(c),  $\nu(T)$ converges at late but finite times to the expected dipolar Chern number $D_y$ across different regimes $m$, as predicted.

\section{Discussion}
In this work, we have introduced a general geometric construction of driven quantum systems called {\it geometric quantum driving}, whose key idea is to use a classical particle to steer a parent quantum Hamiltonian over a choice of trajectory and manifold. 
This results in a multitude of time-dependent systems with structured temporal profiles and physics dependent on the local geometry and global topology of the manifold. 
In particular, we demonstrated the construction of two new classes of quantum drives: hyperbolically-driven quantum systems (HDQS) and nonorientably-driven quantum systems (NDQS), with the choices of geodesic flows on the compact hyperbolic Bolza surface and nonorientable Klein bottle surface respectively. 
We showed how these systems exhibit the physics of measurable, robust, topologically quantized dynamical responses under the conditions of slow driving and being fully-gapped, enabled by classical ergodicity of the geodesic trajectories over the manifolds underlying their construction. 

%
We note these topological responses are reminiscent of a charge pump like the Thouless pump \cite{PhysRevB.27.6083} or energy pump between drives in a quasiperiodically driven system~\cite{martin2017topological,PhysRevB.99.064306,PhysRevLett.125.100601, 
 PhysRevLett.125.160505, PhysRevLett.126.106805}, which   can both be interpreted as a quantum Hall response in the extended frequency space picture~\cite{PhysRevA.7.2203} of Floquet or quasiperiodically-driven systems respectively.
 Indeed, for the Klein bottle quantum drive, we have explained that the topological invariant extracted in dynamics has a direct physical interpretation as the 
 crystalline and electromagnetic response in an effective topological insulator system with unusual nonsymmorphic momentum space symmetries, induced by artificial gauge or translational gauge fields \cite{chen2022brillouin,5nsr-rw4d}.
%
For the HDQS,  we explain  in the SI~\cite{SM} how  their dynamics can  similarly be mapped into a single-particle hopping problem defined on an nonstandard generalized `frequency' lattice governed by a quasienergy operator~\cite{jauslin1991spectral, blekher1992floquet, viennot2018schrodinger}, though the interpretation of their quantized topological response as a crystalline or electromagnetic response in this lattice is not immediately obvious. 
Nevertheless, what these examples show is how geometric quantum driving may be thought of as a useful tool to quantum simulate solid state phenomena realized by nonstandard, exotic lattices. 

More generally, our geometric construction is very versatile and one may consider other trajectories $x_t$ (e.g.~horocycle~\cite{10.1215/S0012-7094-84-05110-X} or Asonov flows~\cite{anosov1967geodesic} etc.), and other manifolds $M$ (e.g.~Klein quartic~\cite{Klein1878}),   giving rise to a plethora of different classes of driven quantum systems.  In the adiabatic limit, and under the assumption of  trajectories chosen to be ergodic, we expect by similar arguments in this work that topological invariants corresponding to surface integrals over the manifold $M$ may be extracted through the dynamical responses of the driven system, using carefully constructed observables.  Moving away from few-body topological physics, it would also be highly interesting to understand these systems' quantum ergodic properties studied via the newly-introduced concept of {\it Hilbert space ergodicity}~\cite{pilatowsky2023complete,PhysRevX.14.041059,dmfd-lgcq}, as well as
what novel dynamical phases of matter can be realized in {\it interacting} systems which are geometrically driven (not necessarily slowly).
Geometric quantum driving thus serves as a powerful bridge linking the topics of quantum dynamics, differential geometry, ergodicity and topology,  and opens new possibilities in quantum simulation. 

\section{Methods}
We provide  details on the hyperbolically-driven qubit model of the main text used in the numerical simulations. 
The parent quantum Hamiltonian is of the form
\begin{align}
H_Q(z)=\bm{d}_\epsilon(z)\cdot \bm{\sigma},
\label{eqn:qubit}
\end{align}
where $\bm{d}_\epsilon(z)=(d_{x,\epsilon}(z),d_{y,\epsilon}(z),d_{z,\epsilon}(z))^T \in \mathbb{R}^3$ is a vector of field strengths defined over the Bolza surface $\Sigma_2$ and  parameterized also by  $\epsilon \in \mathbb{R}$.  The vector $\bm{\sigma} = (\sigma^x,\sigma^y,\sigma^z)$ are the standard Pauli matrices. The vector $\bm{d}_\epsilon(z)$ is defined as follows.


First we introduce a bump function whose domain is the fundamental octagon of the Bolza surface as depicted in Fig.~\ref{fig:4}(a): 
\begin{align}
    f_\rho(z) :=
    \begin{cases}
    A\exp(-\frac{1}{\rho^2-|z|^2})+B, \quad & |z| < \rho \\
    B,\quad & |z| \geq \rho
    \end{cases}
    \label{eqn:bump}
\end{align}
where $A=\pi e^{1/\rho^2}$, $B=-\frac{\pi}{2}$ and $\rho$ is the radius of the bump which we take to be $0.6$ in numerical simulations. This function behaves at such: when $z=0$, $f =\frac{\pi}{2}$. Moving away radially, it smoothly changes to $f = -\frac{\pi}{2}$ at  $|z| = \rho$ and thereafter stays constant in the rest of the Bolza surface. That is to say, the function changes only in a compact region $|z| < \rho$. Note that by construction, $f_{\rho}(z)$ is everywhere smooth (even at the boundary of the fundamental region of the Bolza).

 Then, we define a primitive (non-normalized) vector $\boldsymbol{d}_\epsilon'$ with tuning parameter $\epsilon \in \mathbb{R}$ as follows:
 \begin{equation}
    \boldsymbol{d}_\epsilon'(z)
    =\begin{pmatrix}
        \cos f(z) (z+z^*)/2|z| 
        \\\cos f(z) (z-z^*)/2i|z|  \\

        \sin{f(z)} 
    \end{pmatrix} + 
    \begin{pmatrix}
    0 \\ 0 \\ \epsilon
    \end{pmatrix},
\end{equation}
which is everywhere smooth (even at the boundaries of the fundamental region of the Bolza) by construction. 
Finally, for $|\epsilon|\neq 1$,  we define our desired normalized vector
\begin{equation}
    \boldsymbol{d}_\epsilon(z)=\boldsymbol{d}_\epsilon'(z)/|\boldsymbol{d}_\epsilon'(z)|,
\end{equation}
which is also everywhere smooth and has unit norm (so that $H_Q(z)$ is fully-gapped). 
Note $(z+z^*)/2|z| = x_1/|z|  = \cos\phi$ and $(z-z^*)/2i|z| = x_2/|z| = \sin\phi$
 where $\phi=\arctan(x_2/x_1)$. Thus near the origin, the (unnormalized) spin vector behaves like $\sim (\cos\phi,\sin\phi,\epsilon)^T$; i.e., radially pointing outwards. An illustration of spin vectors  for various $\epsilon$ is shown in Fig.~\ref{fig:4}(a). 
 
 
Now we discuss the topological nature of the spin texture (which is also the topology of the Hamiltonian and its bands).  
At the center of the Bolza surface $|z|=0$, the non-normalized vector $\boldsymbol{d}_\epsilon'$ is
\begin{equation}
    \boldsymbol{d}_\epsilon'(0)
    =\begin{pmatrix}
        0 \\
        0 \\
        1 
    \end{pmatrix} + 
    \begin{pmatrix}
    0 \\ 0 \\ \epsilon
    \end{pmatrix}=\begin{pmatrix}
        0 \\
        0 \\
        1+\epsilon 
    \end{pmatrix}
\end{equation}
while outside the bump $|z|\geq\rho$, the non-normalized vector $\boldsymbol{d}'_\epsilon$ is
\begin{equation}
    \boldsymbol{d}_\epsilon'(0)
    =\begin{pmatrix}
        0 \\
        0 \\
        -1 
    \end{pmatrix} + 
    \begin{pmatrix}
    0 \\ 0 \\ \epsilon
    \end{pmatrix}=\begin{pmatrix}
        0 \\
        0 \\
        -1+\epsilon 
    \end{pmatrix}.
\end{equation}
We see that when $|\epsilon|<1$, the  vector $\boldsymbol{d}_\epsilon$ smoothly changes   from up to down moving radially outward from the center of the Poincar\'e disk, indicating an inversion of the spin texture. We expect this texture to be characterized by a nontrivial topological number, and indeed, in main text, demonstrated how this is captured by a unit Chern number of one of the instantaneous bands of $H_Q(z)$ (up to sign). Since this is a qubit system, we note the Chern number is also equal to the skyrmion number of $\bm{d}_{\epsilon}(z)$. Conversely, when $|\epsilon| >1$, the spin vector $\boldsymbol{d}_\epsilon$ points predominantly in one direction everywhere in $\Sigma_2$ (either up or down), indicating a ferromagnetic texture. We thus expect this texture to harbor a trivial topological number, and indeed, the Chern number of the instantaneous band of $H_Q(z)$ (or skyrmion number of $\bm{d}_\epsilon(z)$) is zero.




%

\section{Data Availability}
The data that support the findings of this study are available from the corresponding author upon reasonable request.

\section{Acknowledgments}
We are grateful to Jiangbin Gong, Ching Hua Lee, Yingcheng Li, Daniel K.~Mark, Hoi Chun Po and Ken Shiozaki for helpful discussions. D.~B.~is supported by the United States Office of Naval Research (N0001425GI01144).
W.~W.~H.~is supported by the National Research Foundation (NRF), Singapore, through the NRF Felllowship NRF-NRFF15-2023-0008, and through the National Quantum Office, hosted in A*STAR, under its Centre for Quantum Technologies Funding Initiative (S24Q2d0009). 

\section{Author Contributions}
D.~B.~and W.~W.~H.~theoretically conceptualized and directed the study. J.~W.~and C.~L.~performed theoretical analyses while J.~W.~performed numerical simulations. All authors discussed the results and contributed to the writing of the manuscript.

\section{Competing Interests statement}
The authors declare no competing interests.

\end{document}


\title{Supplemental Information for  \\
Geometric quantum drives and topological dynamical responses:
hyperbolically-driven quantum systems and beyond}

\author{Jihong Wu}
\affiliation{Department of Physics, National University of Singapore, Singapore 117551}

\author{Chuan Liu}
\affiliation{Department of Physics, National University of Singapore, Singapore 117551}

\author{Daniel Bulmash}
\affiliation{Department of Physics, United States Naval Academy, Annapolis MD 21402 USA}

\author{Wen Wei Ho}
\email{wenweiho@nus.edu.sg}
\affiliation{Department of Physics, National University of Singapore, Singapore 117551}
\affiliation{Centre for Quantum Technologies, National University of Singapore, 3 Science Drive 2, Singapore 117543}

\maketitle

In this supplemental information, we provide details of the geometry of the Bolza surface, as well as calculations supporting our claim of distinct topologically quantized dynamical responses of fully-gapped hyperbolically-driven and nonorientably-driven quantum systems in the limit of slow driving.
Specifically, in Sec.~\ref{sec:Bolza} we introduce the Bolza surface and discuss the nature of geodesics on it, in particular explaining their ergodic nature due to the Hopf theorem.
In Sec.~\ref{sec:derivation} and \ref{sec:derivation2}, we derive our claim of the topologically quantized dynamical response of fully-gapped hyperbolically-driven quantum systems and Klein bottle driven quantum systems respectively. In Sec.~\ref{sec:rp2}, we introduce another nonorientably driven system: a geometric quantum drive constructed from the choice of the real projective plane, and numerically show its novel topological response related to quadrupole Chern number. In Sec.~\ref{sec:counterdiabatic}, we explain how the use of counterdiabatic driving which suppresses diabatic transitions between bands of the parent quantum Hamiltonian, results in quantized dynamical responses without the need for slow driving.
In Sec.~\ref{sec:extended}, we explain how the dynamics of a hyperbolically-driven quantum system can be mapped into the dynamics of a single-particle hopping problem on a generalized `frequency' lattice, governed by a quasienergy operator.

\section{Geometry and geodesics of the Bolza surface}
\label{sec:Bolza}

\subsection{Construction of the Bolza surface}

In this section, we describe the \emph{Bolza surface}  \cite{3603e0c0-3eab-3525-9b0a-caac90c781bb}, a compact, boundaryless Riemann surface of genus $2$ with constant negative curvature. This is the manifold used in the main text  to construct a specific class of  geometric quantum drives, namely hyperbolically-driven quantum systems. While in the literature the symbol $\Sigma_2$ typically refers to any genus $2$ surface in the topological sense, we will refer to $\Sigma_2$ as the particular geometrical surface that is the Bolza.


We begin by considering the {Poincar\'{e} disk model} of the hyperbolic plane $\mathbb{H}$, which consists of the set of points $|z|<1$ with $z := x^1+ix^2$ ($x^i \in \mathbb{R}$) and equipped with the metric
\begin{equation}
    ds^2 = 4\frac{(dx^1)^2+(dx^2)^2}{(1-|z|^2)^2}. 
\end{equation}
%
Every possible orientation-preserving transformation of the Poincar\'e disk to itself, which preserves hyperbolic distances, is given by a fractional linear transformation
\begin{equation}
    z\mapsto Mz, \qquad M:= \frac{az+b}{b^*z+a^*},
\end{equation}
with $a,b \in \mathbb{C}$ and $|a|^2-|b|^2=1$, i.e.~$M$  is an element of the group $\rm SU(1,1)$. Since $M = \pm \mathbb{I}$ represent the same transformation, we may identify the group of orientation-preserving isometries as  $\rm PSU(1,1) = \rm SU(1,1)/\{ \pm \mathbb{I} \}$. 

Now, a finite-volume hyperbolic surface may be constructed by choosing a discrete subgroup $\Gamma$ of $\rm PSU(1,1)$, which is called a Fuchsian group \cite{fuchsian}, and forming the quotient space $\mathbb{H}/\Gamma$. We will use the following Fuchsian group: it is defined by generators $\gamma_1,\gamma_2,\gamma_3,\gamma_4 \in \rm PSU(1,1)$, with
\begin{equation}
    \gamma_1=\frac{1}{\sqrt{1-\sigma^2}}\begin{pmatrix}
        1&\sigma\\\sigma&1
    \end{pmatrix},
\end{equation}
and where $\sigma=\sqrt{2\cos\alpha(1+\cos\alpha)}$ and $\alpha=\pi/4$, so that
\begin{equation}
    \gamma_\mu=
    \begin{pmatrix}
        e^{i(\mu-1)\alpha/2}&0\\0&e^{-i(\mu-1)\alpha/2}
    \end{pmatrix}\gamma_1
    \begin{pmatrix}
        e^{-i(\mu-1)\alpha/2}&0\\0&e^{i(\mu-1)\alpha/2}
    \end{pmatrix},\quad \mu=1,\dots,4.
    \label{eqn:fuchsian}
\end{equation}
The generators of $\Gamma$ obey the relation
\begin{equation}
    \gamma_1\gamma_2^{-1}\gamma_3\gamma_4^{-1}\gamma_1^{-1}\gamma_2\gamma_3^{-1}\gamma_4=\mathbb{I},
\end{equation}
where the inverse transformations $\gamma_\mu^{-1}$ correspond to the matrix inverses of Eq.~\ref{eqn:fuchsian}. 

\begin{figure}[t]
    \centering
    \includegraphics[width=0.4\linewidth]{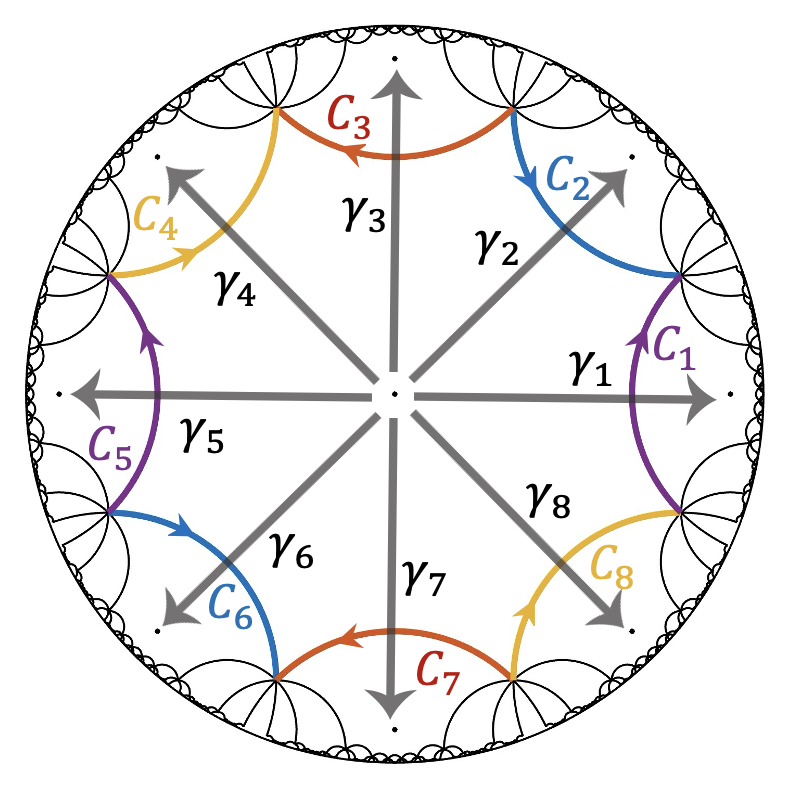}
    \caption{The Bolza surface is the quotient space $\mathbb{H}/\Gamma$, where $\mathbb{H}$ is the hyperbolic plane and $\Gamma$ is a Fuchsian group, here defined by generators $\gamma_1,\dots,\gamma_4$ (translates $\gamma_5,\cdots,\gamma_8$ are their inverses respectively).  Within the Poincar\'e disk model we can identify the fundamental region as the central octagon with its opposite boundary segments $C_j$ identified in the orientation of the arrows shown.}
    \label{fig:Fuchsian}
\end{figure}

On the Poincar\'e disk, this results in a so called $\{8,8\}$ tessellation of hyperbolic space (the notation refers to a tessellation by regular $8$-gons such that each vertex has coordinate number $8$) where each octagon is related to another by some Fuchsian translate, see Supplementary Fig.~\ref{fig:Fuchsian} (note, the translates are non-commuting). Identifying the polygon in the center of the Poincar\'e disk as the fundamental region, the Bolza surface $\Sigma_2$ may be understood as the central octagon with sides glued in the orientation as shown in Supplementary Fig.~\ref{fig:Fuchsian}. 
Its boundary segments $C_j$ can be parameterized as 
\begin{equation}
    C_j = \{z = e^{i(j-1)\frac{\pi}{4}}(c+re^{i\theta}),\frac{3\pi}{4}\leq\theta\leq\frac{5\pi}{4}\},\quad j = 1,\ldots,8,
\end{equation}
which describe arcs of circles, where
\begin{equation}
    c = \sqrt{\frac{3+2\sqrt{2}}{2+2\sqrt{2}}},\quad r = \frac{1}{2+2\sqrt{2}}.
\end{equation}

\subsection{Geodesics on the Bolza surface
}

There are many equivalent formulations of geodesics on a Riemannian manifold $M$ equipped with a metric $g_{ij}(x)$. One such formulation is that of so-called cogeodesic flows generated by the Hamiltonian $H_{cl}(x,p) = \frac{1}{2} g^{ij}(x) p_ip_j$ of a   unit-mass classical particle, freely moving on the manifold ($g^{ij}(x)$ is the inverse metric) \cite{geodesics}. Here $(x,p)$ are local coordinates of the phase space, or in more precise mathematical terms, the   cotangent bundle $T^*M$ of the manifold $M$. 

We apply this formalism to derive geodesics on the hyperbolic plane $\mathbb{H}$, concretely within the Poincar\'e disk model.
Geodesics on the Bolza surface $\Sigma_2$ are then simply obtained by applying the appropriate Fuchsian translate to ensure the trajectories always lie within the fundamental region in the Poincar\'e disk. 
We  first discuss coordinates parameterizing the phase space $T^*\mathbb{H}$  of the Poincar\'e disk. It is given by a pair of complex numbers $(z,p):=(x^1+ix^2,p_1+ip_2)$, where we have conveniently grouped the position coordinates into a complex variable $z := x^1+ix^2$, as well as momentum coordinates into another complex variable $p := p_1 + i p_2$. The metric on the Poincar\'e disk reads
\begin{align}
g_{ij}(z) = g(z) \delta_{ij}, \qquad g(z) = \frac{4}{(1-|z|^2)^2},
\label{Eq:classical motion}
\end{align}
where $i,j = 1,2$. The equations of motion of a unit mass classical particle moving freely are
%
\begin{align}
\dot{x}^i & = \frac{\partial H_{cl}}{\partial p_i} = g^{ij}(z) p_j, \\
\dot{p}_i &= -\frac{\partial H_{cl}}{\partial x^i} = -\frac{1}{2} \frac{\partial g^{jk}(z) }{\partial x^i} p_j p_k,
\end{align}
where Einstein summation convention has been utilized. Here $g^{ij}(z)$ is the inverse metric   $g^{ij}(z)=g(z)^{-1}\delta^{ij}$.
We can compactly write the equations of motion  as
\begin{equation}
(\dot{z},\dot{p}) = \left(\frac{1}{4}(1-|z|^2)^2p,\frac{1}{2}z(1-|z|^2)|p|^2\right)
\label{Eq:flow}
\end{equation}
where $|p|^2=p_1^2+p_2^2$. 
Upon specifying an initial condition $(z_0,p_0)$, there will be a unique solution $(z_t, p_t)$. 
Note that since the system is conservative,  the energy $E = \frac{1}{2}g^{ij}(z_t) p_{i,t} p_{j,t}$ is a constant along the geodesic; equivalently  the magnitude of the momentum (squared) $g^{ij}(z_t) p_{i,t} p_{j,t}$ is conserved.


As mentioned in the main text, given a geodesic $(z_t, p_t)$, we may define a family of related geodesics parameterized by $\lambda > 0$ describing the same path on the Bolza surface via
\begin{align}
(z^{(\lambda)}_t,p_t^{(\lambda)}) = (z_{\lambda t}, \lambda p_{\lambda t}).
\end{align}
These then differ only by their conserved energy $E(\lambda) = \frac{1}{2}g^{ij}(z_0) p_{i,0} p_{j,0} \lambda^2 = E(1) \lambda^2$; thus $\lambda$ may be understood as the speed of the geodesic. We adopt the convention that $E(1) = 1/2$ thus $\lambda = 1$ corresponds to what we refer to as a `unit-speed' geodesic.

An example of a unit-speed geodesic is 
\begin{equation}
    (z_t^{(1)},p_t^{(1)}) = (n\tanh(t/2),n(1+\cosh t))
\end{equation}
where $n=e^{i\theta}$ for some $\theta \in [0,2 \pi)$. This describes a geodesic starting at the origin at $t = 0$ moving in the direction $n$.
By invoking a M\"{o}bius transformation $R(z)=\frac{z+z_0}{1+z z_0^*}$ for some $|z_0|<1$ we can get a new unit-speed trajectory
${z'}^{(1)}_t: = R(z_t^{(1)})=\frac{z_t^{(1)}+z_0}{1+ z_t^{(1)}z_0^*}$ 
which begins at the point $z_0$ at $t = 0$.

\subsection{Ergodicity of geodesics on the Bolza surface}
\label{section:ergodicity}

An important property of geodesic (or cogeodesic) flows on the Bolza surface, which underpins the key results in our work, is the fact that they are {\it ergodic}. That is to say, the only invariant set  under the geodesic flow on the Bolza surface (which we note preserves the standard Liouville measure of phase space) is either the null set or the entire space, up to sets of measure 0. This is guaranteed by a result of Hopf's~\cite{hopf1936fuchsian,bams/1183533160,ergodic}: \\
{\bf Theorem 1 [Hopf 1936]}. 
{\it The geodesic flow on a closed, negatively curved {\it surface} is ergodic.}

This was subsequently improved by Anosov~\cite{anosov1967geodesic}: \\
{\bf Theorem 2 [Anosvov 1967]}. 
{\it The geodesic flow for any compact manifold of negative sectional curvature is ergodic.}



The implication, utilizing Birkhoff's theorem~\cite{doi:10.1073/pnas.17.2.656}, is that if we have an integrable function $f(z,p)$ on the cotangent bundle $T^*\Sigma$ of the Bolza surface, then that for almost all initial conditions $(z_0^{(\lambda)},p_0^{(\lambda)})$ yielding a geodesic $(z_t^{(\lambda)},p_t^{(\lambda)})$, we have
\begin{align}
\lim_{T \to \infty}
\frac{1}{T}\int_0^T \!\! dtf(z_t^{(\lambda)},p_t^{(\lambda)}) = \iint_{\Sigma_2} \!\!\!\!\! dx^1 dx^2\frac{g(z)}{4\pi} \int_0^{2\pi} \!\!\! \frac{d\theta}{2\pi}f(z,\lambda g(z)^{\frac{1}{2}}e^{i\theta}).
\end{align}
In other words, almost surely, time-averaging along a geodesic equals space-averaging (over the level set of $T^*\Sigma_2$ with constant energy $E = \lambda^2/2$). Note the factor $4 \pi$ is the total area of the Bolza surface (computed with the Poincar\'e metric).

Let us numerically illustrate the ergodicity of the geodesic flow on the Bolza. We consider  a small circle with radius $r=0.6$   placed at the center of the Poincar\'e disk, which is completely contained within the fundamental region of the Bolza surface, see Supplementary Fig.~\ref{fig:ergodicity}(a). We define an indicator function $I(z)$ which takes value $1$ within the circle and $0$ otherwise. 
Then, the area of the small circle $S_c=\frac{4\pi r^2}{1-r^2}|_{r=0.6}=\frac{4\pi(0.6)^2}{1-(0.6)^2} \approx 7.06858$, should be estimable through via the ratio of time that a generic geodesic spends within the circle  to the total time elapsed. 
In Supplementary Fig.~\ref{fig:ergodicity}(a) we show the geodesic trajectory considered (red line), which is specified by initial condition $z_0 = 0$ and $p_0 =e^{i \pi/9}$. 
In Supplementary Fig.~\ref{fig:ergodicity}(b) we compute the finite-time estimate of the area
\begin{align}
S_{c,est.}(T) = \frac{4\pi}{T}\int_0^T dt I(z_t). 
\label{eqn:Sest}
\end{align}
We see that $S_{c,est.}(T)$ approaches $S_c$ as $T$ increases, as expected from ergodicity. In the numerical simulation, a precision of 2000 digits is used to guarantee the precision of the trajectory as it traverses the Bolza surface (we chose this large precision as the motion on the Bolza is known to be chaotic).



\begin{figure}[t]
    \centering
    \includegraphics[width=0.9\linewidth]{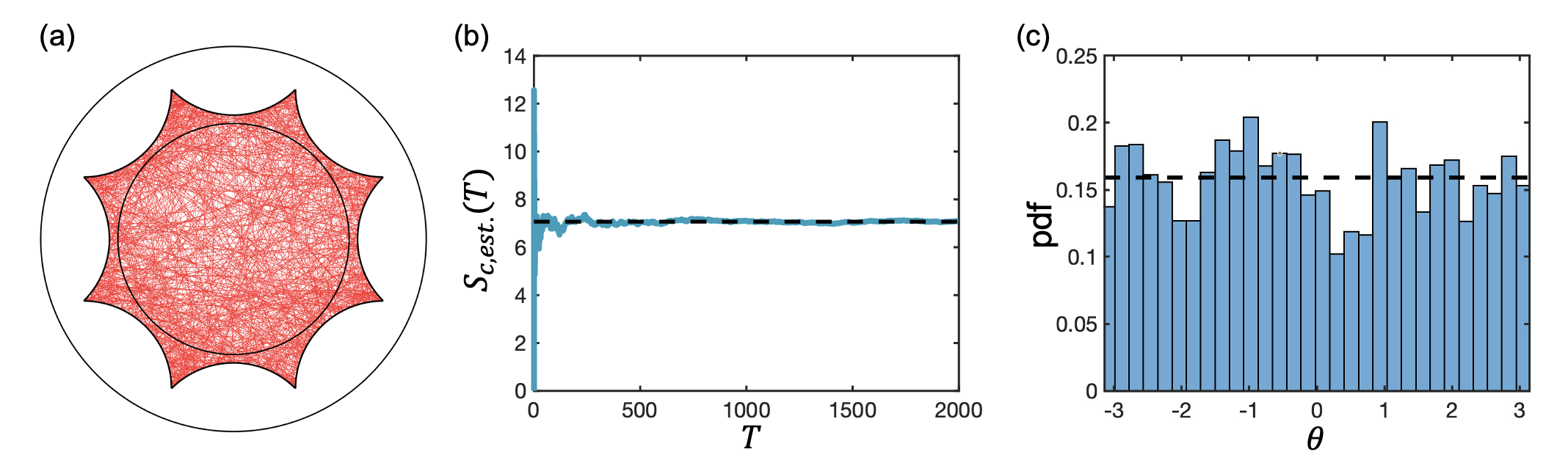}
    \caption{(a) A  circle of radius $r= 0.6$ is placed at the center of the Bolza surface. The red line corresponds to a geodesic specified by initial condition $z_0 = 0$ and $p_0 =e^{i \pi/9}$ evolved up to $t=2000$. (b) Finite-time approximation of the area of the circle using Eq.~\eqref{eqn:Sest}. 
    As $T$ increases, $S_{c,est.}(T)$  converges to the area of the circle $S_c$ (black dashed line). (c) Probability distribution function of $\theta$. Black dashed line represents the uniform distribution.}
    \label{fig:ergodicity}
\end{figure}

We also numerically illustrate  ergodicity in the momentum variable. We consider the angle $\theta\equiv\arctan(p_y/p_x)$ and empirically obtain the distribution of $\theta$ along the geodesic flow $z_t$ considered above, whenever the flow enters the circle $|z_t| < 0.6$. 
The probability distribution function of $\theta$ is shown in Supplementary Fig.~\ref{fig:ergodicity}(c), which is seen to be close to the expected uniform distribution over the interval $[-\pi,\pi]$. 


\section{Derivation of the topologically quantized response of fully-gapped hyperbolically-driven quantum systems in the limit of slow driving}
\label{sec:derivation}

In this section, we derive our claim made in the main text that fully-gapped hyperbolically-driven quantum systems in the adiabatic limit exhibit a topologically quantized dynamical response.

We remind the reader of our set-up and claim: given a fully-gapped $D$-level quantum Hamiltonian $H_Q(z)$ and a choice of a geodesic $(z_t^{(\lambda)}, p_t^{(\lambda)})$, we initialize the system at $t=0$ in the instantaneous eigenstate of the $n$-th band $|\psi_n(z_0^{(\lambda)})\rangle$ at the point $z_0^{(\lambda)}$. 
Then, we drive the system using the time-dependent quantum Hamiltonian $H(t) := H_Q(z_t^{(\lambda)})$ to obtain the time-evolved quantum state $|\psi(t)\rangle$.
We then measure the $D$-level observable
\begin{equation}
    O^{(\lambda)}(t):= 2g(z_t^{(\lambda)})^{-1}\left(p^{(\lambda)}_{2,t} \partial_{x^1}H_Q(z^{(\lambda)}_{t})-p^{(\lambda)}_{1,t} \partial_{x^2}H_Q(z^{(\lambda)}_{t})\right),
    \label{eqn:O}
\end{equation}
and  time-average it (with rescaling too), to obtain
\begin{align}
w(T) := \frac{1}{\lambda^2 T} \int_0^T dt \langle \psi(t)| O^{(\lambda)}(t)|\psi(t)\rangle.
\end{align}
Our claim is that 
\begin{align}
\lim_{\lambda \to 0} \lim_{T\to \infty} w(T) = C_1^{(n)},
\label{eqn:main_claim}
\end{align}
where 
\begin{align}
C_1^{(n)} = \frac{1}{2\pi}\iint_{\Sigma_2} dx^1 dx^2 \Omega_{x^1 x^2}(z)
\end{align}
is the first Chern number of the $n$-th band. Here $\Omega_{x^1 x^2}^{(n)}$ is its Berry curvature form
\begin{align}
\Omega^{(n)}_{x^1 x^2}(z) := \partial_{x^2}A^{(n)}_{x^1}(z) - \partial_{x^1} A^{(n)}_{x^2}(z),
\end{align}
where $A^{(n)}_{x^j} = i \langle \psi_n(z)|\partial_{x^j} \psi_n(z)\rangle$ is the Berry connection.


\subsection{Derivation}
Our derivation consists of a computation of   $w(T)$ performed within an adiabatic series expansion up to first-order, where $\lambda$ (the speed of the drive) is the relevant small parameter. The fact that we utilize an adiabatic series expansion is natural, since by assumption that the system is fully-gapped, the speed $\lambda$ can always be taken to be much smaller than the energy gap between the $n$ and $n \pm 1$ bands; then, we can expect the system stays close to the instantaneous eigenstate $|\psi_n(z_t^{(\lambda)})\rangle$ for long times. However, importantly, our calculation will show that the first order corrections --- namely, the diabatic transitions to other eigenstates $|\psi_m(z_t^{(\lambda)}\rangle$ ($m \neq n$) --- are essential to deriving the quantized response.

We begin by expanding the time-evolved quantum state $|\psi(t)\rangle$ in an adiabatic series in $\lambda$. Following  standard formalism (see for example Eq.~10.95 of {\it Introduction to Quantum Mechanics, Second Edition}, David J.~Griffiths~\cite{adiabatic}, applied to our scenario, we have the expansion to first order
\begin{align}
|\psi(t)\rangle  =  e^{i \phi_n(t)} e^{i \gamma_n(t)}|\psi_n(z_t^{(\lambda)})\rangle   - \sum_{m\neq n} e^{i \phi_m(t)} |\psi_m(z_t^{(\lambda)})\rangle \left( \int_0^t ds \langle \psi_m(z_s^{(\lambda)})|\partial_s \psi_n(z_s^{(\lambda)})\rangle e^{i \gamma_n(s)} e^{i (\phi_n(s) - \phi_m(s))} \right),
\label{eqn:correction}
\end{align}
where the dynamic phase $\phi_m$ and Berry phase $\gamma_m$ are given by 
\begin{align}
\phi_m(t) &\equiv -\int_0^t ds E_m(z_s^{(\lambda)}), \\
\gamma_m(t) & \equiv i \int_0^t ds \langle \psi_m(z_s^{(\lambda)})|\partial_s \psi_m(z_s^{(\lambda)})\rangle. 
\end{align}
It is important to note that the time-dependence of the state $|\psi(t)\rangle$ on the LHS of Eq.~\eqref{eqn:correction} is {\it generated} by the Schr\"odinger equation, while the time-dependence of objects like the states  $|\psi_m(z_t^{(\lambda)})\rangle$ and energies $E_m(z_t^{(\lambda)})$ on the RHS arise from the {\it evaluation} of the instantaneous eigenstates $|\psi_n(z)\rangle$ along the geodesic $z_t^{(\lambda)}$.



Since all objects on the RHS which harbor the time derivative $\partial_s$ have time-dependencies arising from evaluation of a parent quantum state along a geodesic, we can rewrite the time derivative as
\begin{align}
\frac{\partial}{\partial s} = g(z_s^{(\lambda)})^{-1} p_{1,s}^{(\lambda)} \frac{\partial}{\partial x^1} + g(z_s^{(\lambda)})^{-1} p_{2,s}^{(\lambda)} \frac{\partial}{\partial x^2} + \left(-\frac{1}{2} \frac{\partial g^{jk}(z_s^{(\lambda)})}{\partial x^1} p_{j,t}^{(\lambda)} p_{k,t}^{(\lambda)}\right) \frac{\partial}{\partial p_1} + \left(-\frac{1}{2} \frac{\partial g^{jk}(z_s^{(\lambda)})}{\partial x^2} p_{j,t}^{(\lambda)} p_{k,t}^{(\lambda)}\right) \frac{\partial}{\partial p_2}.
\label{eqn:timederivative}
\end{align}


This allows us to write the time-evolved state Eq.~\eqref{eqn:correction} as 
\begin{align}
|\psi(t)\rangle  = e^{i \phi_n(t)} e^{i \gamma_n(t)}|\psi_n(z_t^{(\lambda)})\rangle - \sum_{m\neq n} e^{i \phi_m(t)} |\psi_m(z_ t^{(\lambda)})\rangle \left( \int_0^t ds g(z_s^{(\lambda)})^{-1}p_{i,s} ^{(\lambda)}  \langle \psi_m(z_s^{(\lambda)})|\partial_{x^i} \psi_n(z_s^{(\lambda)})\rangle e^{i \gamma_n(s)} e^{i (\phi_n(s) - \phi_m(s))} \right),
\label{eqn:correction2}
\end{align}
where we used the fact that  $\partial_{ p_i}|\psi_n(z_t^{(\lambda)})\rangle$ vanishes. 

We now evaluate the expectation value of  $O^{(\lambda)}(t)$ using the state Eq.~\eqref{eqn:correction2}, which can itself be expanded in an adiabatic series:
\begin{align}
\langle \psi(t)|O^{(\lambda)}(t) |\psi(t)\rangle = o^{(\lambda)}_1 + o_2^{(\lambda)}+O(\lambda^3).
\end{align}
Note the $a$-th term $o_a^{(\lambda)}$ is of order $\sim \lambda^a$. The reason the series starts from $a=1$ instead of $a = 0$ is because the observable $O^{(\lambda)}(t)$ itself is $\lambda$-small, carrying a factor of momentum $p^{(\lambda)}_{i,t}$. 

 The first order term  $o_1^{(\lambda)}$ comes from the expectation  value of $O^{(\lambda)}(t)$ evaluated in the first term  of the expansion of the time-evolved state Eq.~\eqref{eqn:correction2}.
Let us for the sake of brevity consider only the first term $2g(z_t^{(\lambda)})^{-1}p_{2,t}^{(\lambda)} \partial_{x^1}H_Q(z_t^{(\lambda)})$ of $O^{(\lambda)}(t)$ (Eq.~\eqref{eqn:O}); the second term entails a similar calculation. Then, its time-average is
\begin{align}
    & \frac{2}{T}\int_0^Tdt\langle\psi_n(z_t^{(\lambda)})|g(z_t^{(\lambda)})^{-1}p_{2,t}^{(\lambda)}\partial_{x^1}H_Q(z_t^{(\lambda)})|\psi_n(z_t^{(\lambda)})\rangle \nonumber \\
    = &\frac{2}{T}\int_0^Tdtg(z_t^{(\lambda)})^{-1}p_{2,t}^{(\lambda)}\partial_{x^1}E_n(z_t^{(\lambda)}) \nonumber \\
    \stackrel{T \to \infty}{=} &2\iint_{\Sigma_2}dx^1dx^2\frac{g(z)}{4\pi}g(z)^{-1}
    \lambda g(z)^{\frac{1}{2}}
    \partial_{x^1}E_n(z)
\int_0^{2\pi}\frac{d\theta}{2\pi}\sin(\theta) \nonumber \\
    = & 0.
\end{align}
The first equality arises because of the Feynman-Hellmann theorem. The second equality is due to  Hopf's argument [{\bf Theorem 1} of Sec.~\ref{section:ergodicity}]. The third equality is because $\int_0^{2\pi} d\theta \sin\theta=0$.

Now, we move on to the second order term $o_2^{(\lambda)}$. This comes from matrix elements of $O^{(\lambda)}(t)$ evaluated between the first and second terms of Eq.~\eqref{eqn:correction2} and its complex conjugate. 
Again let us consider only  the first term $2g(z_t^{(\lambda)})^{-1}p_{2,t}^{(\lambda)} \partial_{x^1}H_Q(z_t^{(\lambda)})$ of $O^{(\lambda)}(t)$ (Eq.~\eqref{eqn:O}) for now.  We have the expression:
\begin{align}
-\frac{2}{T} \int_0^T dt \sum_{m\neq n} \langle \psi_m(z_t^{(\lambda)})| g(z_t^{(\lambda)})^{-1}p_{2,t}^{(\lambda)}\partial_{x^1}H_Q(z_t^{(\lambda)})|\psi_n(z_t^{(\lambda)})\rangle e^{i (\phi_n(t) - \phi_m(t)) } e^{i \gamma_n(t)} G(t,\lambda) + \text{c.c.},
\end{align}
where  
\begin{equation}
    G(t,\lambda)=\int_0^t ds g(z_s^{(\lambda)})^{-1}p_{i,s}^{(\lambda)}   \langle \partial_{x^i} \psi_n(z_s^{(\lambda)})|\psi_m(z_s^{(\lambda)})\rangle e^{-i \gamma_n(s)} e^{-i (\phi_n(s) - \phi_m(s))}.
    \label{eqn:Gt}
\end{equation}

We use the fact that, for $m \neq n$,
\begin{align}
\langle \psi_m(z_t^{(\lambda)})|\partial_{x^1} H_Q(z_t^{(\lambda)})|\psi_n(z_t^{(\lambda)})\rangle = (E_n(z_t^{(\lambda)}) - E_m(z_t^{(\lambda)})) \langle \psi_m(z_t^{(\lambda)})|\partial_{x^1}\psi_n(z_t^{(\lambda)})\rangle
\end{align}
and
\begin{align}
  i \partial_t \left( e^{i(\phi_n(t) - \phi_m(t))} \right) =   (E_n(z_t^{(\lambda)})-E_m(z_t^{(\lambda)}))e^{i (\phi_n(t) - \phi_m(t))},
\end{align}
so we can recast the expression as
\begin{align}\label{eq:S32}
-\frac{2i}{T}  \sum_{m\neq n}  \int_0^T dt 
\underbrace{\partial_t \left( e^{i (\phi_n(t) - \phi_m(t))}   \right)}_{u'(t)}
\underbrace{
\langle \psi_m(z_t^{(\lambda)})|\partial_{x^1} \psi_n(z_t^{(\lambda)})\rangle  e^{i \gamma_n(t)}g(z_t^{(\lambda)})^{-1}p_{2,t}^{(\lambda)} G(t,\lambda)}_{v(t)} +\text{c.c.}. 
\end{align}
Above, we can have written the integral as $-\frac{2i}{T}\sum_{m \neq n} \int_0^T dt u' (t)v(t)$ which invites us to perform  integration by parts:
\begin{align}
-\frac{2i}{T}\sum_{m \neq n} \int_0^T dt u' (t)v(t) = -\frac{2i}{T}\sum_{m\neq }[u(t)v(t)]_0^T + \frac{2i}{T}\sum_{m \neq n}\int_0^T dt u(t)v'(t).
\end{align}
The boundary term (ignoring the complex conjugate) is
\begin{align}
-\frac{2i}{T}\sum_{m \neq n} [u(t) v(t)]_0^T,
\label{eqn:boundary}
\end{align}
and we want to estimate its magnitude. Now, $u(t) = e^{i(\phi_n(t) - \phi_m(t))}$ is a $U(1)$ phase factor so has unit norm, while all factors of $v(t)$ except $G(t,\lambda)$ have bounded norm by virtue of the Hamiltonian being smooth and the manifold being compact. As for $G(t,\lambda)$, we note it is the leading order correction to the adiabatic expansion in Eq.~\eqref{eqn:correction2} so is $\lambda$ small by the adiabatic theorem. We further observe that its maximum  value appears {\it uniformly} bounded from above  in time, see  Supplementary Fig.~\ref{fig:Gt_norm}. Assuming this behavior persists to infinite times, then Eq.~\eqref{eqn:boundary} vanishes as $T \to \infty$ (for fixed $\lambda$).

We are then only left with the term $\frac{2i}{T}\sum_{m \neq n}\int_0^T dt u(t)v'(t) + \text{c.c.}$, which reads
\begin{align}
 \frac{2i}{T} \sum_{m \neq n} \int_0^T dt e^{i (\phi_n(t) - \phi_m(t))} \partial_t \left( 
\langle \psi_m(z_t^{(\lambda)})|\partial_{x^1} \psi_n(z_t^{(\lambda)})\rangle  e^{i \gamma_n(t)}g(z_t^{(\lambda)})^{-1}p_{2,t}^{(\lambda)} G(t,\lambda)\right) + \text{c.c.}. 
\label{eqn:uvprime}
\end{align}
Let us first differentiate in time   the term $G(t,\lambda)$. We end up with one contribution of the form
\begin{align}
\frac{2i}{T} \sum_{m \neq n} \int_0^T dt (g(z_t^{(\lambda)})^{-1})^2p_{2,t}^{(\lambda)}p_{i,t}^{(\lambda)}\langle \psi_m(z_t^{(\lambda)})|\partial_{x^1}\psi_n(z_t^{(\lambda)})\rangle \langle \partial_{x^i} \psi_n(z_t^{(\lambda)})|\psi_m(z_t^{(\lambda)})\rangle + \text{c.c.}.
\label{eqn:Chern_term}
\end{align}
Since $\langle \psi_n|\partial_{x^i} \psi_n\rangle$ is purely imaginary,  we can `complete' the sum $\sum_{m \neq n} \to \sum_{m}$ (i.e., lift the restriction). This is because the inclusion of the $m=n$ term adds nothing to Eq.~\eqref{eqn:Chern_term} (it and its complex conjugate cancel).
Then, we can get rid of this sum entirely, changing the matrix element to $\langle \partial_{x^i} \psi_n|\partial_{x^1} \psi_n\rangle$. Since $\langle \partial_{x^1}\psi_n | \partial_{x^1} \psi_n\rangle$ is real, this matrix element also vanishes (again, because $i \langle \partial_{x^1}\psi_n | \partial_{x^1} \psi_n\rangle + \text{c.c} = 0$). As a result, we have a simplification to
\begin{align}
\frac{2i}{T}  \int_0^T dt (g(z_t^{(\lambda)})^{-1})^2p_{2,t}^{(\lambda)}p_{2,t}^{(\lambda)}\langle \partial_{x^2} \psi_n(z_t^{(\lambda)})|\partial_{x^1}\psi_n(z_t^{(\lambda)})\rangle + \text{c.c.}.
\end{align}
In the limit of large $T$ we then again invoke ergodicity [{\bf Theorem 1} of Sec.~\ref{section:ergodicity}] of the geodesic trajectory  to 
%
change the time integral to a phase space integral, which yields
\begin{align} 
& \frac{i}{2\pi}\iint_{\Sigma_2} g(z)dx_1dx_2\int_0^{2\pi}d\theta g(z)^{-1}\frac{\lambda^2\sin^2\theta}{2\pi}\langle \partial_{x^2} \psi_n(z)|\partial_{x^1}\psi_n(z)\rangle+\text{c.c.}. \nonumber \\
= &\frac{i\lambda^2}{4\pi}\iint_{\Sigma_2} dx_1dx_2\langle \partial_{x^2} \psi_n(z)|\partial_{x^1}\psi_n(z)\rangle-\langle\partial_{x^1} \psi_n(z)|\partial_{x^2}\psi_n(z)\rangle
\nonumber \\
= &\frac{\lambda^2C_1^{(n)}}{2}.
\label{eqn:s39}
\end{align}

We still have to apply the time derivative on the other factor $\langle \psi_m (z_t^{(\lambda)})|\partial_{x^1} \psi_n(z_t^{(\lambda)}) \rangle  e^{i \gamma_n(t)}g(z_t^{(\lambda)})^{-1}p_{2,t}^{(\lambda)}$ in Eq.~\eqref{eqn:uvprime}. This yields
\begin{align}
\frac{2i}{T} \sum_{m \neq n} \int_0^T dt e^{i (\phi_n(t) - \phi_m(t)} \partial_t\left( \langle \psi_m (z_t^{(\lambda)})|\partial_{x^1} \psi_n(z_t^{(\lambda)}) \rangle  e^{i \gamma_n(t)}g(z_t^{(\lambda)})^{-1}p_{2,t} ^{(\lambda)}\right) 
G(t,\lambda).
\label{eqn:remaining_term}
\end{align}
But we can replace $\partial_t$ by the derivative along the geodesic Eq.~\eqref{eqn:timederivative} (changing $s \to t$). 
Estimating the size of Eq.~\eqref{eqn:remaining_term} we find it is of order $\lambda^3$ (recall $G(t,\lambda)$ is of order $\lambda$) and is thus sub-leading to the Chern number contribution Eq.~\ref{eqn:Chern_term}. 

Finally, let us not forget the contribution from the second part of the operator $O^{(\lambda)}(t)$, $-2g(z_t^{(\lambda)})^{-1}p_{1,t}^{(\lambda)} \partial_{x^2}H_Q(z_t^{(\lambda)})$. However, an identical computation (e.g.~changing $\sin^2\theta \to \cos^2\theta$ in Eq.~\eqref{eqn:s39}) yields also 
the contribution
$\frac{\lambda^2C_1^{(n)}}{2}$. Combining both contributions, we then obtain our main claim Eq.~\eqref{eqn:main_claim}.
\begin{figure}[t]
    \centering
    \includegraphics[width=0.9\linewidth]{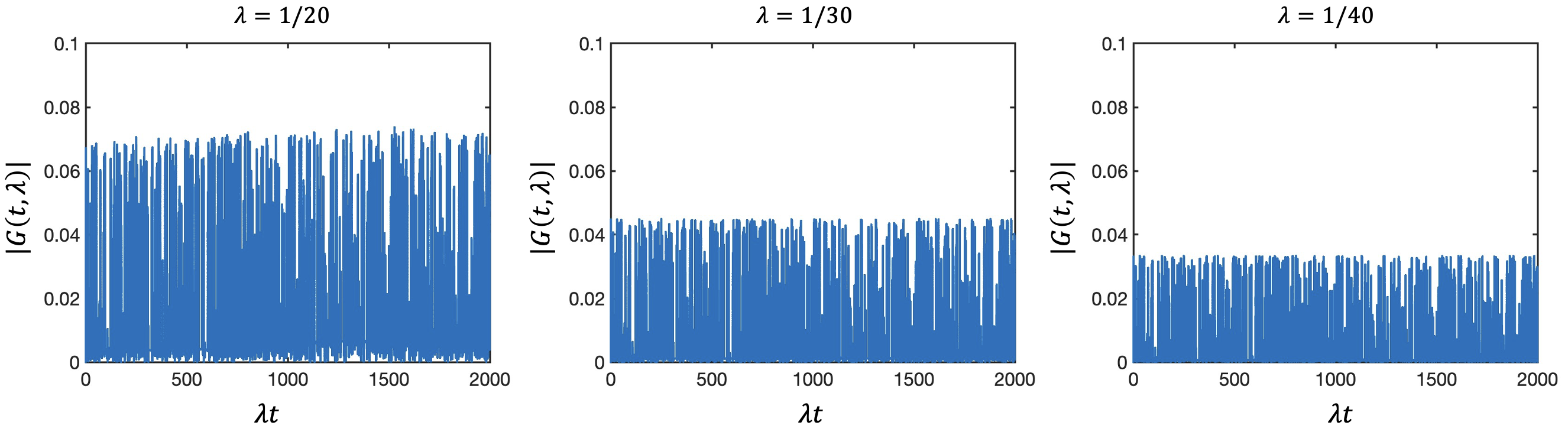}
    \caption{Plots of $|G(t,\lambda)|$ [Eq.~\ref{eqn:Gt}] versus $t$ for the hyperbolically-driven qubit model numerically simulated in the main text, for various $\lambda$s.
    One sees that    $|G(t,\lambda)|$ is of order $\lambda$, and is uniformly bounded from above for the range of times simulated. 
    }
    \label{fig:Gt_norm}
\end{figure}


\section{Derivation of the topologically quantized response of fully-gapped nonorientably-driven quantum systems in the limit of slow driving}
\label{sec:derivation2}
The derivation proceeds similarly to that in the previous section, with a minor modification to the time derivative. For quasiperiodic driving, the parameter dependence takes the form $\bm\theta_t=\bm\omega t+\bm\theta_0$, so the time derivative can be expressed as $\partial_t=\omega_i\partial_{\theta_i}$, where $\omega_i$ denotes a constant velocity on the torus. However, for Klein bottle driving the velocity along the $\theta_x$ direction is no longer constant. Instead, it reverses sign whenever the trajectory crosses $\theta_y=n\pi$ with $n\in\mathbb{Z}$, reflecting the nonorientable identification of the Klein bottle. As a result, the time derivative is modified to
\begin{equation}
    \frac{\partial}{\partial_t}=(-1)^{n_y(t)}\omega_x\frac{\partial}{\partial_{\theta_x}}+\omega_y\frac{\partial}{\partial_{\theta_y}}
\end{equation}
where $\omega_y$ remains constant and the $x$ direction velocity becomes time-dependent and piecewise constant. Here, $n_y(t)=\lfloor(\omega_yt+\theta_{y,0})/\pi\rfloor$ counting the number of crossings of $\theta_y=n\pi$ up to time $t$. For notational convenience, we denote the time-dependent velocity $(-1)^{n_y(t)}\omega_x$ by $\omega_x(t)$ in the following.

We also evaluate the expectation value $\langle\psi(t)|O^{(\boldsymbol{\omega})}|\psi(t)\rangle$ to leading order in $\|\boldsymbol{\omega}\|$ at the adiabatic limit, where $O^{(\boldsymbol{\omega})}(t)=\omega_y\theta_{y}(t)\frac{\partial H_Q(\boldsymbol{\theta}_t)}{\partial\theta_x}$ and $|\psi(t)\rangle$ is a time evolving state starting from an eigenstate of the initial Hamiltonian. The first order term is expectation value of the instantaneous eigenstates $|\psi_n(\boldsymbol{\theta}_t)\rangle$  :
\begin{align}
    &\frac{1}{T}\int_0^Tdt\langle \psi_n(\boldsymbol{\theta}_t)|\omega_y\theta_{y}(t)\frac{\partial H_Q(\boldsymbol{\theta}_t)}{\partial\theta_x}|\psi_n(\boldsymbol{\theta}_t)\rangle\nonumber\\
    =&\frac{\omega_y}{T}\int_0^Tdt\theta_{y}(t)\frac{\partial E_n(\boldsymbol{\theta}_t)}{\partial\theta_x}\nonumber\\
    \overset{T\to\infty}{=}&\frac{\omega_y}{2\pi^2}\int_{-\pi}^0 d\theta_y\int_0^{2\pi}d\theta_x\theta_y\frac{\partial E_n(\boldsymbol{\theta})}{\partial\theta_x}\nonumber\\=&0.
\end{align}
The second equality arises because of ergodicity of the classical trajectory on the parameter space, which allows the time average to be replaced by a uniform average over the fundamental domain. The final equality hold because $E_n(\bm \theta)$ is periodic in $\theta_x$, implying $E_n(2\pi,\theta_y)=E_n(0,\theta_y)$.

We now turn to the second order term. The expansion is similar to that of the previous section, where Eq.~$\eqref{eq:S32}$ is modified as
\begin{equation}
    -\frac{i}{T}  \sum_{m\neq n}  \int_0^T dt 
\underbrace{\partial_t \left( e^{i (\phi_n(t) - \phi_m(t))}   \right)}_{u'(t)}
\underbrace{
\langle \psi_m(\boldsymbol{\theta}_t)|\partial_{\theta_x} \psi_n(\boldsymbol{\theta}_t))\rangle  e^{i \gamma_n(t)}\omega_y\theta_{y}(t)G(t,\boldsymbol{\omega})}_{v(t)} +\text{c.c.}. 
\end{equation}
where
\begin{equation}
    G(t,\boldsymbol{\omega})=\int_0^t ds \omega_i(s)   \langle \partial_{\theta_i} \psi_n(\boldsymbol{\theta}_t)|\psi_m(\boldsymbol{\theta}_t)\rangle e^{-i \gamma_n(s)} e^{-i (\phi_n(s) - \phi_m(s))}.
\end{equation}
After integrating by parts, we retain only the leading contribution, $\frac{2i}{T}\sum_{m \neq n}\int_0^T dt u(t)v'(t) + \text{c.c.}$ and within $v'(t)$, we keep only the derivative of $G(t,\boldsymbol{\omega})$, which yields,
\begin{align}
    \frac{i}{T}\sum_{m\neq n}\int_0^Tdt\omega_y\omega_i(t)\theta_{y}(t)\langle\psi_m(\boldsymbol{\theta}_t)|\partial_{\theta_x}\psi_n(\boldsymbol{\theta}_t)\rangle\langle\partial_{\theta_i}\psi_n(\boldsymbol{\theta}_t)|\psi_m(\boldsymbol{\theta}_t)\rangle+\text{c.c.}.
\end{align}
This can be simplified by replacing $\sum_{m\neq n}$ with $\sum_m$ and discarding the term $\langle\partial_{\theta_1}\psi_n(\boldsymbol{\theta})|\partial_{\theta_1}\psi_n(\boldsymbol{\theta})\rangle$, yielding
\begin{align}
    &\frac{i}{T}\sum_{m\neq n}\int_0^Tdt\omega_y^2\theta_{y}(t)\langle\partial_{\theta_y}\psi_n(\boldsymbol{\theta}_t)|\partial_{\theta_x}\psi_n(\boldsymbol{\theta}_t)+\text{c.c.}\nonumber\\
    \overset{T\to\infty}{=}&\frac{\omega_y^2}{2\pi^2}\iint_{K^2}d\theta_xd\theta_y\theta_y\Omega^{(n)}(\boldsymbol{\theta})\nonumber\\
    =&\frac{\omega_y^2}{\pi}D_y^{(n)}.
\end{align}
So we have shown that in the long-time and slow driving limit, the response $\nu(T)$ converges corresponding to the quantized dipolar Chern number:
\begin{align}
    \lim_{\|\boldsymbol{\omega}\|\to0}\lim_{T\to\infty}\frac{\pi}{\omega_y^2T}\int_0^Tdt\langle\psi(t)|O^{(\boldsymbol{\omega})}(t)|\psi(t)\rangle=D_y^{(n)}.
\end{align}

\section{Topological response of the real projective plane driven models}\label{sec:rp2}
In this section, we introduce another nonorientably driven model, in which the underlying manifold is chosen to be the real projective plane $RP^2$. As in the Klein bottle driven case, imposing an additional symmetry on the parent Hamiltonian gives rise to a quantized topological response. In contrast to the dipolar Chern response associated with Klein bottle driving, driving on $RP^2$ leads to a distinct invariant: the quadrupolar Chern number, which characterizes the fully gapped phases of the system in the adiabatic limit.

For the real projective space, it can be constructed as the quotient $\mathbb{R}^2/\chi$, where $\chi$ is generated by 
\begin{equation}
    \left\{\begin{aligned}
        \chi_1&:(x,y)\mapsto (-x+\pi,y+\pi),\\
        \chi_2&:(x,y)\mapsto(x+\pi,-y+\pi).
    \end{aligned}
    \right.
\end{equation}
Hence, the fundamental region for $RP^2$ can be chosen to be the square $ \mathcal{F}=[0,\pi]\times[0,\pi]$ (gray region in Supplementary Fig.~\ref{fig:rp2}(a)). Since $\chi_1^2$ and $\chi_2^2$ correspond exactly two translations on the torus, four copies of the real projective plane can be combined to form a torus, as illustrated in Supplementary Fig.~\ref{fig:rp2}(a) (white and gray regions). With the flat metric inherited from $\mathbb{R}^2$, geodesics on the covering space are straight lines, considering the particle's initial velocity $\bm\omega=(\omega_x,\omega_y)$ and initial position $\bm\theta_0=(\theta_{x,0},\theta_{y,0})$, then
\begin{equation}
\tilde{\boldsymbol{\theta}}(t)=\boldsymbol{\theta}_0+\boldsymbol{\omega}\,t,
\qquad
\tilde{\theta}_x(t)=\theta_{x,0}+\omega_x t,\quad
\tilde{\theta}_y(t)=\theta_{y,0}+\omega_y t,
\end{equation}
and a geodesic on $RP^2$ is obtained by projecting $\tilde{\boldsymbol{\theta}}(t)$ back to $\mathcal{F}$ using the group action. The generators implement orientation-reversing identifications: whenever the lifted trajectory crosses a $\pi$-cell boundary in the $y$-direction one applies $\chi_1$, which wraps $y$ back into $[0,\pi)$ while reflecting $x\mapsto \pi-x$; similarly, each $\pi$-cell crossed in the $x$-direction triggers $\chi_2$, which wraps $x$ while reflecting $y\mapsto \pi-y$. 
Introducing the crossing numbers
\begin{equation}
n_x(t)\equiv \left\lfloor {\tilde{\theta}_x(t)}/{\pi}\right\rfloor,
\qquad 
n_y(t)\equiv \left\lfloor {\tilde{\theta}_y(t)}/{\pi}\right\rfloor,
\end{equation}
and the wrapped coordinates
\begin{equation}
\bar{\theta}_x(t)\equiv \tilde{\theta}_x(t)\bmod \pi,\qquad 
\bar{\theta}_y(t)\equiv \tilde{\theta}_y(t)\bmod \pi,
\end{equation}
so that $\bar{\theta}_{x,y}(t)\in[0,\pi)$, the projected geodesic $\boldsymbol{\theta}(t)\in\mathcal{F}$ can be written compactly as
\begin{equation}
\theta_x(t)=(-1)^{n_y(t)}\bar{\theta}_x(t)+\frac{\pi}{2}\big[1-(-1)^{n_y(t)}\big],\qquad
\theta_y(t)=(-1)^{n_x(t)}\bar{\theta}_y(t)+\frac{\pi}{2}\big[1-(-1)^{n_x(t)}\big].
\end{equation}
Equivalently, $\theta_x=\bar{\theta}_x$ ($\pi-\bar{\theta}_x$) for even ( odd) $n_y$, and $\theta_y=\bar{\theta}_y$ ($\pi-\bar{\theta}_y$) for even (odd) $n_x$. Thus, while the lifted motion has constant velocity $\boldsymbol{\omega}$, the projected velocity on $RP^2$ undergoes sign flips whenever $n_x$ or $n_y$ changes, reflecting the nonorientable identifications encoded by $\chi_1$ and $\chi_2$. Explicitly, the resulting time dependent velocities are
\begin{equation}
    \omega_x(t)=(-1)^{n_y(t)}\omega_x,\quad\omega_y(t)=(-1)^{n_x(t)}\omega_y.
\end{equation}


Consider now a parent Hamiltonian on the real projective plane $H_Q({\bm\theta})$ and ${\bm\theta}\in[0,\pi]\times[0,\pi]$. 
Because $RP^2$ is nonorientable, the conventional first Chern number does not exist. 
However, suppose we impose the following symmetry:
\begin{equation}
        S:H_Q(\theta_x,\theta_y)=U^\dagger H_Q(\theta_y,-\theta_x+\pi)U
    \end{equation}
where $U$ is a unitary operator satisfying $U^4=\pm I$. With the $S$ symmetry, the $n$-th eigenstates $|\psi_n(\boldsymbol{\theta})\rangle$ (assuming it is gapped) {\it are} now topologically classified by a quantized Berry curvature quadrupole moment, called the quadruple Chern number \cite{5nsr-rw4d}, defined as such:
\begin{equation}
    Q_{xy}^{(n)}=\frac{1}{\pi}\int_{0}^\pi\int_0^\pi d\theta_xd\theta_y\theta_x\theta_y\Omega^{(n)}(\theta_x,\theta_y),
\end{equation}
where $\Omega^{(n)}(\bm\theta)$ is the Berry curvature of the $n$-th band. In the context of crystalline and electromagnetic responses in topological insulators in solid state physics, artificial gauge or translational gauge fields can effectively realize the real projective plane Brillouin zones \cite{5nsr-rw4d}. Further, it was shown that the integer $Q_{xy}^{(n)}$ can be interpreted as the rate of momentum density bound to translationally symmetric lattice defects, such as homogeneous strains or dislocation lines, or a momentum current generated by the time derivative of strain and shear deformations.

In a fully gapped $RP^2$ quantum drive, this exotic topological invariant can be extracted using an analogous protocol as in the HDQS and  Klein bottle drive, but by measuring the following time-dependent operator:
\begin{equation}
    O^{(\boldsymbol{\omega
    })}(t):=\omega_{y}(t)\theta_{x}(t)\theta_{y}(t)\frac{\partial H_Q(\bm\theta_t)}{\partial\theta_x}.
\end{equation}
Notice that $\omega_{y}(t)$ is also time-dependent here. Then compute its time-averaged expectation value:
\begin{equation}
    \mu(T)=\frac{\pi}{\omega_{y}(t)^2T}\int_0^Tdt\langle\psi(t)|O^{(\boldsymbol{\omega
    })}(t)|\psi(t)\rangle.
\end{equation}
In the long-time and slow driving limit, we claim the response $\mu(T)$ converges corresponding to the quantized quadrupolar Chern number:
\begin{equation}
    \lim_{\|\bm\omega\|\to0}\lim_{T\to\infty}\mu(T)=Q_{xy}^{(n)}.
\end{equation}
The derivation proceeds analogously as before, with the understanding that geodesics on $RP^2$ are ergodic. 

To illustrate this physics, we performed numerical simulation of a qubit $RP^2$ quantum drive. We consider a drive with parent quantum Hamiltonian
\begin{equation}
    H_Q(\bm\theta)=\bm d(\bm\theta)\cdot\bm\sigma,
\end{equation}
where $\bm\sigma$ are the standard Pauli matrices and  $\bm d(\bm\theta)$ is chosen to be
\begin{equation}\label{eq:rp2}
    \begin{aligned}
        &d_x(\bm\theta)=\sin\theta_x\sin2\theta_y,\\
        &d_y(\bm\theta)=\sin2\theta_x\sin\theta_y,\\
        &d_z(\bm\theta)=m+\cos2\theta_x+\cos2\theta_y.
    \end{aligned}
\end{equation}
Above $m$ is a tunable parameter, and the symmetry operator $U$ can be seen to be $\exp(i\pi\sigma_z/2)$, motivated by~\cite{5nsr-rw4d}.
For generic $m$ the Hamiltonian is fully gapped, with a topologically nontrivial regime occuring for $0<m<2$, where $Q_{xy}^{(n)}=\pi^2/2$. In contrast, the trivial regime occurs for $m>2$ with $Q_{xy}^{(n)}=0$.
To implement the drive, we consider a geodesic $\bm\theta_t$ on  $RP^2$ defined by $\omega_x=0.02$ and $\omega_y=\frac{1+\sqrt{5}}{2}\omega_x$ starting at $\bm\theta = \bm{0}$. 
A temporal profile of the model with $m=1$ is shown in Supplementary Fig.~\ref{fig:rp2}(b).
Using this, we numerically simulate the system beginning from an instantaneous eigenstate of the parent quantum Hamiltonian at $t = 0$, measure $O^{(\bm\omega)}(t)$, before computing $\mu(T)$. The   results are shown in Supplementary Fig.~\ref{fig:rp2}(c)(d), which demonstrates that $\mu(T)$ approaches the quadrupole Chern number for both trivial and topological regimes at late times, as expected.

\begin{figure}[t]
    \centering
    \includegraphics[width=1\linewidth]{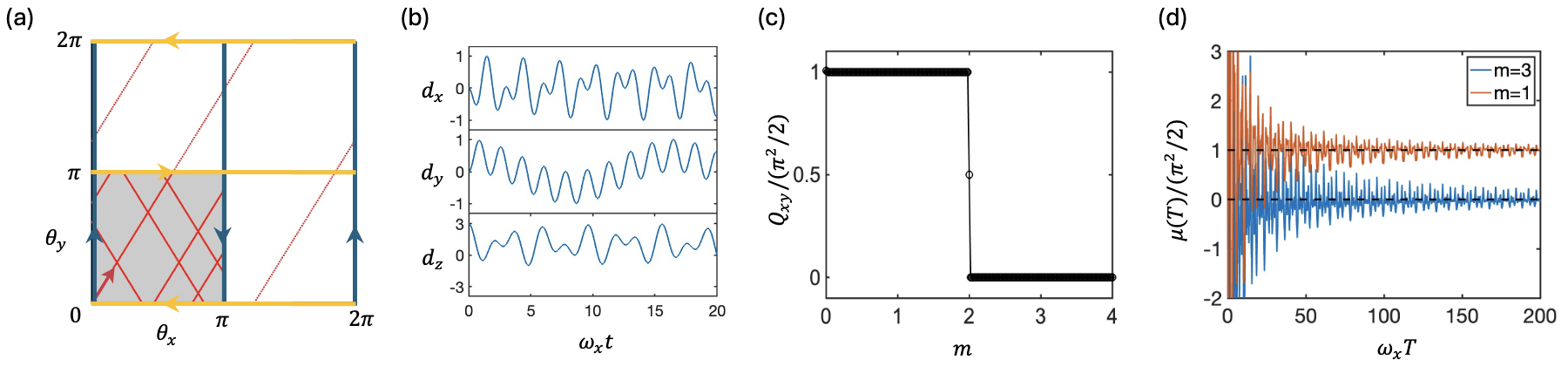}
    \caption{(a) Geodesics on the real projective plane (gray region) are depicted by red straight lines. Every time the particle leaves the gray region, it gets mapped back to a point within the fundamental region following the identification rules of the boundaries (yellow and blue arrows).  
    Equivalently, one may consider four copies of the real projective planes glued as shown (white and gray regions) which yield a standard torus; a real projective plane geodesic is then equivalent  to a standard geodesic on the torus (dotted trajectory). (b) Temporal profile of the example $RP^2$ driven qubit Hamiltonian  with $m = 1$. (c) The quadrupole Chern number of the parent quantum Hamiltonian Eq.~\eqref{eq:rp2}. $Q_{xy}$ is quantized to $\pi^2/2$ in the regime $0<m<2$, and becomes topologically trivial $(Q_{xy}=0)$ when $m>2$. (d) Dynamical response $\mu(T)$ starting from the instantaneous eigenstate of the upper band. For a small driving frequency as stated in the main text, $\mu(T)$ approaches to the quadrupole Chern number of the band at late times, in both trivial and topological regimes.}
    \label{fig:rp2}
\end{figure}

\section{Counterdiabatic driving and topologically quantized response without the need for slow driving}
\label{sec:counterdiabatic}

In the main text, we discussed the topological quantized response of a fully-gapped hyperbolically-driven quantum system in the slow driving limit. 
Here, we explain how we can lift the requirement of slow driving, if we utilize counterdiabatic (CD) driving  which suppresses diabatic transitions between instantaneous eigenstates of $H_Q(z_t^{(\lambda)})$.

\subsection{Counterdiabatic Hamiltonian and claim}


We follow the exposition of Ref.~\cite{Berry_2009}, which explains that to keep quantum dynamics completely within the instantaneous eigenstate $|\psi_n(t)\rangle$ of a time-dependent Hamiltonian $H(t)$   (beginning at $t=0$ with the state $|\psi_n(0)\rangle$), one can employ counterdiabatic driving which entails adding an additional term $V(t)$ to $H(t)$, to wit
\begin{equation}
    H_{CD}(t) = H(t)+V(t) = H(t)+ \sum_{m\neq n}i \frac{|\psi_m(t)\rangle\langle \psi_m(t)|\partial_tH(t)|\psi_n(t)\rangle\langle \psi_n(t)|}{E_n(t)-E_m(t)}.
    \label{Eq:CD general}
\end{equation}

We employ this framework to our hyperbolically-driven quantum 
Hamiltonian $H(t) = H_Q(z_t^{(\lambda)})$ where $(z_t^{(\lambda)},p_t^{(\lambda)})$ is a geodesic on the Bolza surface $\Sigma_2$. The CD Hamiltonian $H_{CD}(t)$ comprises $H_Q(z_t^{(\lambda)})$ together with 
\begin{align}
    V(t) & =  \sum_{m\neq n}i \frac{|\psi_m(z_t^{(\lambda)})\rangle\langle \psi_m(z_t^{(\lambda)})|\partial_tH_Q(z_t^{(\lambda)})|\psi_n(z_t^{(\lambda)})\rangle\langle \psi_n(z_t^{(\lambda)})|}{E_n(z_t^{(\lambda)})-E_m(z_t^{(\lambda)})} \nonumber \\
    & = \dot{x}^{(\lambda),i}(t) \sum_{m\neq n}i \frac{|\psi_m(z_t^{(\lambda)})\rangle\langle \psi_m(z_t^{(\lambda)})|\partial_{x^i}H_Q(z_t^{(\lambda)})|\psi_n(z_t^{(\lambda)})\rangle\langle \psi_n(z_t^{(\lambda)})|}{E_n(z_t^{(\lambda)})-E_m(z_t^{(\lambda)})} \nonumber \\
    & = p_{i,t}^{(\lambda)} g(z_t^{(\lambda)})^{-1}   \sum_{m\neq n}i \frac{|\psi_m(z_t^{(\lambda)})\rangle\langle \psi_m(z_t^{(\lambda)})|\partial_{x^i}H_Q(z_t^{(\lambda)})|\psi_n(z_t)\rangle\langle \psi_n(z_t^{(\lambda)})|}{E_n(z_t^{(\lambda)})-E_m(z_t^{(\lambda)})}.
    \label{eqn:V}
\end{align}


An important observation  is that $V(t)$ can be obtained by evaluating  a parent quantum Hamiltonian $V_Q(z,p)$ defined on the cotangent bundle $T^*\Sigma_2$
along the geodesic $(z_t^{(\lambda)},p_t^{(\lambda)})$. That is to say, if we define
\begin{equation}
    V_Q(z,p) = \sum_{m\neq n}i p_i g^{-1}(z)  \frac{|\psi_m(z)\rangle\langle \psi_m(z)|\partial_{x^i}H_Q|\psi_n(z)\rangle\langle \psi_n(z)|}{E_n(z)-E_m(z)},
\end{equation}
then we have
\begin{align}
V(t) = V_Q(z,p)|_{(z,p)=(z_t^{(\lambda)},p_t^{(\lambda)})}.
\end{align}
This also means that the counterdiabatic Hamiltonian itself is expressible as the evaluation of a parent Hamiltonian along the geodesic, namely
\begin{align}
H_{CD}(t) = H_{CD,Q}(z,p)|_{(z,p)=(z_t^{(\lambda)},p_t^{(\lambda)})}
\end{align}
where 
\begin{align}
H_{CD,Q}(z,p) = H_Q(z) + V_Q(z,p)
\end{align}
is its parent Hamiltonian.

Now, the observable we will measure is Eq.~\eqref{eqn:O} but with the replacement $H_Q(z) \mapsto H_{CD,Q}(z,p)$, namely
\begin{align}
O_{CD}^{(\lambda)}(t)= 2g(z^{(\lambda)}_t)^{-1} p_{2,t}^{(\lambda)}\partial_{x^1} H_{CD,Q}(z_t^{(
\lambda)},p_t^{(\lambda)}) - 2g(z^{(\lambda)}_t)^{-1} p_{1,t}^{(\lambda)}\partial_{x^2} H_{CD,Q}(z_t^{(
\lambda)},p_t^{(\lambda)})
\label{eqn:OCD}
\end{align}
Our claim is that beginning from the state $|\psi_n(z_0^{(\lambda)})\rangle$ at $t = 0$, with dynamics generated by the counterdiabatic Hamiltonian $H_{CD}(t)$ to obtain the time-evolved state $|\psi(t)\rangle$, the time-averaged expectation value of $O_{CD}^{(\lambda)}(t)$ is quantized in the long time limit: 
\begin{align}
\lim_{T \to \infty} w_{CD}(T) = C_1^{(n)},
\label{eqn:wcd_claim}
\end{align}
where $w_{CD}(T) := \frac{1}{\lambda^2T}\int_0^Tdt\langle \psi(t)|O^{(\lambda)}_{CD}(t)|\psi(t)\rangle$.
Importantly, compared to Eq.~\eqref{eqn:main_claim}, there is no need for the additional limit $\lambda \to 0$.

\subsection{Proof} 
We compute the time-averaged expectation value of the observable Eq.~\eqref{eqn:OCD} within the state $|\psi(t)\rangle$, which by construction is exactly the instantaneous band evaluated along the geodesic: $|\psi(t)\rangle \propto|\psi_n(z_t^{(\lambda)})\rangle$.
For brevity, we consider only the expectation value of the first term of the observable Eq.~\eqref{eqn:OCD} for now. 
%
We get:
\begin{align}
 & \frac{2}{T}\int_0^T dt \langle  \psi_n(z_t^{(\lambda)})|g(z^{(\lambda)}_t)^{-1}p^{(\lambda)}_{2,t} \partial_{x^1}H_{CD,Q}(z^{(\lambda)}_{t},p_t^{(\lambda)}) | \psi_n(z_t^{(\lambda)}) \rangle  \nonumber \\ 
= & \frac{2}{T}\int_0^T dt  g(z_t^{(\lambda)})^{-1}p^{(\lambda)}_{2,t} \left({\rm Tr}|\psi_n(z^{(\lambda)}_t)\rangle\langle \psi_n(z^{(\lambda)}_t)|\partial_{x^1}\left( H_Q(z_t^{(\lambda)}) +V_Q(z^{(\lambda)}_t,p_t^{(\lambda)})\right)\right).
\label{eqn:expr1}
\end{align}
Now consider the long-time limit. Hopf's ergodic theorem allows us to convert the time-integral to a phase space integral, but the term proportional to $\partial_{x^1}H_Q(z_t^{(\lambda)})$ has a single factor of $p_{2,t}^{(\lambda)}$ attached to it and thus averages to 0. 
Therefore, only the term proportional to $\partial_{x^1}V_Q(z_t^{(\lambda)}, p_t^{(\lambda)})$ survives.
Furthermore, we observe 
$\partial_{x^1}V_Q(z_t^{(\lambda)}, p_t^{(\lambda)})$ is the sum of two terms  each of which is proportional to the momentum $p_{1,t}^{(\lambda)}$ and $p_{2,t}^{(\lambda)}$. 
Noting that Eq.~\eqref{eqn:expr1} carries the factor $p_{2,t}^{(
\lambda)}$ too, then the time-average of $p_{1,t}^{(\lambda)} p_{2,t}^{(\lambda)}$ vanishes too, leaving only the term $(p_{2,t}^{(\lambda)})^2$. Therefore,  Eq.~\eqref{eqn:expr1}  further evaluates to
\begin{align}
&   \iint_{\Sigma_2}  dx^1 dx^2\frac{g(z)}{2\pi} \int_0^{2\pi}\frac{d\theta}{2\pi}\lambda^2\sin^2\theta\  {\rm Tr}\left(|\psi_n(z)\rangle\langle \psi_n(z)|\partial_{x^1}\left(\sum_{m\neq l}i g(z)^{-1} \frac{|\psi_m(z)\rangle\langle \psi_m(z)|\partial_{x^2}H_Q(z)|\psi_l(z)\rangle\langle \psi_l(z)|}{E_l(z)-E_m(z)}\right)\right) \nonumber \\
= & i\lambda^2\iint_{\Sigma_2} \frac{dx^1 dx^2}{4\pi}{\rm Tr}\left(|\psi_n(z)\rangle\langle \psi_n(z)|\partial_{x^1} \left(\sum_{m \neq l} |\psi_m(z)\rangle\langle \psi_m(z)|\partial_{x^2}\psi_l(z)\rangle\langle \psi_l(z)| \right)\right) \nonumber \\
=& i\lambda^2\iint_{\Sigma_2} \frac{dx^1 dx^2}{4\pi}{\rm Tr}\left(|\psi_n(z)\rangle\langle \psi_n(z)| \left(\sum_{m \neq l} |\partial_{x^1}\psi_m(z)\rangle\langle \psi_m(z)|\partial_{x^2}\psi_l(z)\rangle\langle \psi_l(z)|+|\psi_m(z)\rangle\langle \psi_m(z)|\partial_{x^2} \psi_l(z)\rangle\langle \partial_{x^1} \psi_l(z)| \right)\right) \nonumber \\
= & i\lambda^2\iint_{\Sigma_2} \frac{dx^1 dx^2}{4\pi}\left(\sum_{m \neq n} \langle \psi_n(z)|\partial_{x^1}\psi_m(z)\rangle\langle \psi_m(z)|\partial_{x^2}\psi_n(z)\rangle+\langle \psi_n(z)|\partial_{x^2} \psi_m(z)\rangle\langle \partial_{x^1} \psi_m(z)|\psi_n(z)\rangle \right),
\label{eqn:expr2}
\end{align}
where we used that $\frac{\langle \psi_m(z)|\partial_{x^2}H_Q(z)|\psi_n(z)\rangle}{E_n(z)-E_m(z)} = \langle \psi_m(z)|\partial_{x^2}\psi_n(z)\rangle$ in the first equality. We further notice that
\begin{align}
    & \sum_{m \neq n} \langle \psi_n(z)|\partial_{x^1}\psi_m(z)\rangle\langle \psi_m(z)|\partial_{x^2}\psi_n(z)\rangle+\langle \psi_n(z)|\partial_{x^2} \psi_m(z)\rangle\langle \partial_{x^1} \psi_m(z)|\psi_n(z)\rangle \nonumber \\
    = &  -\langle \partial_{x^1}\psi_n(z)|\partial_{x^2}\psi_n(z)\rangle+\langle \partial_{x^2}\psi_n(z)|\partial_{x^1}\psi_n(z)\rangle
\end{align}
due to the fact that $\langle\partial_{x^i} \psi_m(z)|\psi_n(z)\rangle + \langle \psi_m(z)|\partial_{x^i} \psi_n(z)\rangle = \partial_{x^i}\langle \psi_m(z)|\psi_n(z)\rangle = 0$ for $m \neq n$. Thus we find that Eq.~\eqref{eqn:expr2} evaluates to
\begin{equation}
    i\iint_{\Sigma_2} \frac{dx^1 dx^2}{4\pi}\left( \langle \partial_{x^2}\psi_n|\partial_{x^1}\psi_n\rangle-\langle \partial_{x^1}\psi_n|\partial_{x^2}\psi_n\rangle\right) = \frac{C_{1}^{(n)}}{2},
\end{equation}
where $C_{1}^{(n)}$ denotes the first Chern number of the $n$-th band of $H_Q(z)$. Similarly,  the second term of Eq.~\eqref{eqn:OCD} evaluates to $\frac{C_{1}^{(n)}}{2}$. Taken together, we have our claim Eq.~\eqref{eqn:wcd_claim}.

\section{Mapping of hyperbolically-driven quantum system to hopping problem on a generalized `frequency' lattice}
\label{sec:extended}

It is possible to map the dynamics of a hyperbolically-driven $D$-level quantum system $H(t) = H_Q(z_t)$ 
acting on a physical Hilbert space $\mathcal{H}$, into a static Hamiltonian problem on an enlarged Hilbert space. The formalism is laid out in Refs.~\cite{jauslin1991spectral,blekher1992floquet,viennot2018schrodinger}. Upon choosing some ordering of basis states in the extended space, one then obtains a hopping problem of a $D$-level particle on a generalized `frequency' lattice, the equivalent of the commonly-used frequency lattice in periodically and quasiperiodically driven quantum systems.

Let us explain how this arises. For simplicity, we assume that the geodesic flow used to construct a hyperbolically-driven quantum system is of unit-speed. Then,  the invariant measure under this ergodic flow is the Liouville measure $d\mu$ on the unit cotangent bundle $UT^*\Sigma_2$, which we describe by local coordinates $(z,p)$. We may treat the parent quantum Hamiltonian $H_Q(z)$ as a function  on the unit cotangent bundle, promoting it to $H_Q(z,p)$ (in an abuse of notation). 
 Now the manifold $UT^*\Sigma_2$ is compact, so we may consider the Hilbert space $L^2(UT^*\Sigma_2,d\mu)$ of complex square-integrable functions on it. 
The enlarged Hilbert space is then $\mathcal{K} := \mathcal{H} \otimes L^2(UT^*\Sigma_2,d\mu)$.

Let us next note that the generator $P$ of geodesic flows is  
\begin{equation}
P(z,p):= -i \left( g^{ij}(z)p_j\partial_{x^i}-\frac{1}{2}\partial_{x^i}g^{jk}(z)p_jp_k\partial_{p_i} \right).
\end{equation}
That is, given a function $f(z,p) \in L^2(UT^*\Sigma_2,d\mu)$,  we have 
\begin{align}
e^{i tP}f(z_0,p_0) = f(z_t,p_t),
\end{align}
such that $(z_t,p_t)$ is  a geodesic with initial condition $(z_0,p_0)$. 
Then, one can define the so-called {\it quasienergy operator} (QEO) $K$, defined to act on the enlarged Hilbert space $\mathcal{K}$, as
\begin{align}
K = H_Q(z,p)+P(z,p),
\end{align}
where $P$'s domain has been enlarged (naturally) to $\mathcal{K}$.
The claim is then that 
\begin{align}
|\psi(t)\rangle := e^{i tP} e^{-itK}|\psi(0)\rangle
\label{eqn:ansatz}
\end{align}
solves the Schr\"odinger equation
\begin{align}
i \partial_t |\psi(t)\rangle = H_Q(z_t)|\psi(t)\rangle,
\end{align}
with initial condition $|\psi(0)\rangle$. To see this is true, we differentiate Eq.~\eqref{eqn:ansatz} to get
\begin{align}
i \partial_t |\psi(t)\rangle & = e^{i t P} H_Q(z,p) e^{-itK}|\psi(0)\rangle \nonumber \\
& = e^{i t P} H_Q(z,p) e^{-i t P} e^{i t P} e^{-i t K} |\psi(0)\rangle \nonumber \\
& = H_Q(z_t,p_t) |\psi(t)\rangle \nonumber \\
& = H_Q(z_t) |\psi(t)\rangle.
\end{align}
Furthermore, it is clear that the initial condition is satisfied. 

The spectral properties  of the QEO  $K$ (i.e.., whether the spectrum is discrete, continuous, or mixed) determine the stability of dynamics of the system (i.e., whether motion is regular or have decaying correlations). For example, suppose $K$ has a discrete spectrum, that is, one can find quasienergies $\xi_n$ and quasienergy states $|\phi_n(z,p)\rangle \in \mathcal{K}$ satisfying
\begin{align}
K |\phi_n(z,p)\rangle = \xi_n |\phi_n(z,p)\rangle.
\end{align}
Then one would be able to expand the dynamics Eq.~\eqref{eqn:ansatz} as 
\begin{align}
|\psi(t)\rangle & = \sum_n e^{-i \xi_n t} e^{i t P} |\phi_n(z,p)\rangle c_n \nonumber \\
& = \sum_n e^{-i \xi_n t} |\phi_n(z_t,p_t)\rangle c_n,
\label{eqn:dynamics_expand}
\end{align}
where the coefficient is extracted via
\begin{align}
c_n = \int_{UT^*\Sigma_2} d\mu \langle \phi_n(z,p)|\psi(0)\rangle.
\end{align}
The astute would immediately recognize the identical formalism of the extended Hilbert space for Floquet systems: there, the eigenstates of the quasienergy operator would be quasienergy states $|\phi_n(\theta)\rangle \in \mathcal{H}\otimes L^2(\theta,d\theta/2\pi)$, which are guaranteed to exist by Floquet's theorem. This implies the dynamics of a periodically-driven quantum system is always decomposable as
\begin{align}
|\psi(t)\rangle = \sum_n e^{-i \xi_n t} |\phi(\omega t)\rangle c_n,
\end{align}
where $c_n = \int_{S^1} \frac{d\theta}{2\pi}\langle \phi_n(\theta)|\psi(0)\rangle$. However, a crucial difference in our present case of a hyperbolically driven quantum system is that there is no {\it guarantee} that normalizable eigenstates and corresponding eigenvalues exist (in fact, this is true even in the class of quasiperiodically-driven quantum systems \cite{jauslin1991spectral,blekher1992floquet}), and thus, no guarantee that their dynamics is decomposable as Eq.~\eqref{eqn:dynamics_expand}. 
Understanding the spectral properties of the quasienergy operator is a nontrivial task, which is beyond the scope of this work.

Nevertheless, one can understand the action of the QEO in terms of a  single-particle hopping model of a fermion with $D$ flavors on a generalized `frequency' lattice, as such. Let $\{ |\sigma\rangle \otimes f_n(z,p) \}_{\sigma=1,\cdots,D; n  \in \mathbb{Z}}$ be a basis for the extended Hilbert space $\mathcal{K}$. We introduce complex fermionic creation operators $\{ c^\dagger_{\sigma,n} \}_{\sigma,n}$ where $n$ represents the site index and $\sigma$ the flavor index (on each site). 
Then $K$ can be mapped onto a single-particle hopping Hamiltonian on the infinitely-many fermionic modes
\begin{align}
K = \sum_{nm\sigma\sigma'} K_{nm}^{\sigma\sigma'}c_{\sigma,n}^\dagger c_{\sigma',m},
\label{eqn:Khop}
\end{align}
where the matrix elements are
\begin{align}
K_{nm}^{\sigma \sigma'}= \int_{UT^*\Sigma_2} d\mu \langle\sigma|K(z,p)|\sigma'\rangle f_n^*(z,p)f_m(z,p).
\end{align}
The utility of this mapping as written, though, is not evident: for example, it is not clear if the model Eq.~\eqref{eqn:Khop} is   local, or what the number of spatial dimensions it lives in is. 

Note crucially that there is a choice of basis $f_n(z,p)$ involved to construct the hopping model, and different choices may lead to different interpretations, some more useful than others. 
For example, let us explain how the standard extended space formalism of a quantum system quasiperiodically-driven  with a $d$-tone frequency vector $\bm{\omega} = (\omega_1, \cdots, \omega_d)$,  as a $d$-dimensional local hopping model, already entailed a particular choice of basis states. Now, the quasienergy operator is
\begin{align}
K = H_Q(\bm\theta) + P(\bm\theta)
\end{align}
which acts in the Hilbert space $\mathcal{K} = \mathcal{H}\otimes L^2(T^d,\prod_i^d\frac{d\theta_i}{2\pi})$, 
where
\begin{align}
H_Q(\bm\theta) = \sum_{n \in \mathbb{Z}^d} H_{\bm n}e^{i \bm{n}\cdot \bm\theta}
\end{align}
and  the generator $P$ of geodesic flows is
\begin{align}
P(\bm\theta) = -i \sum_{i=1}^d \omega_i\partial_{\theta^i}.
\end{align}
Applying the general construction as explained above, upon considering a basis for $\mathcal{K}$ as $\{ |\sigma\rangle \otimes f_n(\bm\theta)\}_{\sigma=1,\cdots,D; n \in \mathbb{Z}}$
we will be able to write down a  fermionic hopping model Eq.~\eqref{eqn:Khop} with matrix elements given by
\begin{align}
K^{\sigma \sigma'}_{nm} = \int_{T^d} \prod_{i=1}^d\frac{d\theta_i} {2\pi} \langle \sigma|K(\bm\theta)|\sigma'\rangle f_n^*(\bm\theta) f_m(\bm\theta).
\end{align}
However, Eq.~\eqref{eqn:Khop}  is  not the usual form that the quasienergy operator of a quasiperiodically driven quantum system is typically found written in the literature. 
The reason is because a standard choice of basis for $\mathcal{K}$   has typically   been implicitly made from the get-go:
\begin{align}
\{ |\sigma\rangle \otimes e^{i \bm{n} \cdot \bm\theta} \}_{\sigma=1,\cdots,D; \bm{n} \in \mathbb{Z}^d}.
\end{align}
That is, instead of some basis states $f_n(\bm\theta)$ of $L^2(T^d,\prod_{i=1}^d\frac{d\theta_i}{2\pi})$ indexed by $n \in \mathbb{Z}$, one  works immediately with the  Fourier basis $e^{i \bm{n} \cdot \bm\theta}$ indexed by a list of $d$ integers $\bm{n} = (n_1,\cdots,n_d) \in\mathbb{Z}^d$ (note that a bijection exists between $\mathbb{Z}$ and $\mathbb{Z}^d$), the latter of which can be interpreted as sites on a $d$-dimensional lattice. This choice is astute, because the Fourier modes are simultaneously both the eigenstates of the translation operator $P$ and also the basis onto which the parent quantum Hamiltonian  $H_Q(\bm\theta)$ is expanded over in a Fourier series, resulting in a translationally-invariant action of $H_Q(\bm\theta)$'s  on the lattice. This results in the familiar expression of the QEO
\begin{align}
K = \sum_{\bm{n}\bm{k}\sigma\sigma'} \langle \sigma|H_{\bm k}|\sigma'\rangle c_{\bm{n}+\bm{k},\sigma}^\dagger c_{\bm{n},\sigma'} + \sum_{\bm{n}\sigma}\bm{\omega}\cdot\bm{n} c^\dagger_{\bm{n},\sigma} c_{\bm{n},\sigma},
\end{align}
which accords the   useful picture of the system as a single $D$-flavored fermion hopping in a square lattice of $d$ spatial dimensions, in the presence of a potential $\bm{\omega}\cdot\bm{n}$ linearly increasing along the direction $\bm{\omega}$, which is incommensurate to the lattice. This leads to immediate physical insights like being able to declare there is Stark localization along the direction $\bm\omega$ etc..
 Furthermore,  the range of hopping in this model can immediately be seen to be tied to the decay of Fourier modes $H_{\bm k}$ with $\bm{k}$, which is itself linked to the smoothness of the parent quantum Hamiltonian $H_Q(\bm\theta)$, by standard Fourier analysis~\cite{steinFourier}.  

Returning   to the hyperbolically-driven quantum system case, one may hope to repeat a similar analysis. However, here it is not clear that eigenstates of the momentum operator $P$ exist, nor what their explicit forms are.
Exploring the different forms of the mapped fermionic hopping models for hyperbolically driven quantum systems, and which form yields useful physical insights, 
would be highly interesting. We leave these investigations for  future work.